\documentclass[useAms,usenatbib]{mn2e}
\usepackage{amsmath} 
\usepackage{graphicx}
\usepackage{tabularx}
\usepackage{lscape,graphicx}
\usepackage{amssymb}
\usepackage[english]{babel}
\usepackage[utf8]{inputenc}
\usepackage{soul}

\usepackage{natbib}
\usepackage{paralist}
\usepackage{aas_macros}

\title{Hubble-Lema\^itre fragmentation and the path to equilibrium of merger-driven cluster formation}

\author[ J. Dorval, C.M. Boily, E. Moraux, T. Maschberger, Ch. Becker]
       { J. Dorval$^1$, C.M. Boily$^{1,2}$, E. Moraux$^{3,4}$, T. Maschberger$^{3,4}$, Ch. Becker$^{3,4}$\\
        $^{1}$Observatoire Astronomique de Strasbourg, Universit\'e de Strasbourg, 11 rue de l'Universit\'e, 67000 Strasbourg,\\
        $^{2}$CNRS UMR 7550, 11 rue de l'Universit\'e, 67000 Strasbourg,\\
        $^{3}$Univ. Grenoble Alpes, IPAG, F-38000 Grenoble, France,\\
		$^{4}$CNRS, IPAG, F-38000 Grenoble, France.}

\AtBeginDocument{}

\begin{document}

\newcommand{\hh}[1]{^{#1}}
\newcommand{\PBaSS}{PBaSS\ }
\newcommand{\cmb}[1]{{\bf #1}}
\newcommand{\inv}[1]{\frac{1}{#1}}

\newcommand{\Hub}{{\cal H}}

\maketitle 

\begin{abstract}
This paper discusses a new method to generate self-coherent initial conditions for young substructured stellar cluster. The expansion of a uniform system allows stellar sub-structures (clumps) to grow from fragmentation modes by adiabatic cooling. We treat the system mass elements as stars, chosen according to a Salpeter mass function, and the time-evolution is performed with a collisional N-body integrator. This procedure allows to create a fully-coherent relation between the clumps' spatial distribution and the underlying velocity field. The cooling is driven by the gravitational field, as in a cosmological Hubble-Lema\^itre flow. The fragmented configuration has a `fractal'-like geometry but with a self-grown velocity field and mass profile. We compare the characteristics of the stellar population in clumps with that obtained from hydrodynamical simulations and find a remarkable correspondence between the two in terms of the stellar content and the degree of spatial mass-segregation. In the fragmented configuration, the IMF power index is $\approx$ 0.3 \textit{lower} in clumps in comparison to the field stellar population, in agreement with observations in the Milky Way. We follow in time the dynamical evolution of fully fragmented and sub-virial configurations, and find a soft collapse, leading rapidly to equilibrium (timescale of 1 Myr for a $\sim 10^4 M_\odot$ system). The low-concentration equilibrium implies that the dynamical evolution including massive stars is less likely to induce direct collisions and the formation of exotic objects. Low-mass stars already ejected from merging clumps are depleted in the end-result stellar clusters, which harbour a top-heavy stellar mass function.

\end{abstract}



\begin{figure*}
\begin{center}
\includegraphics[width=\textwidth]{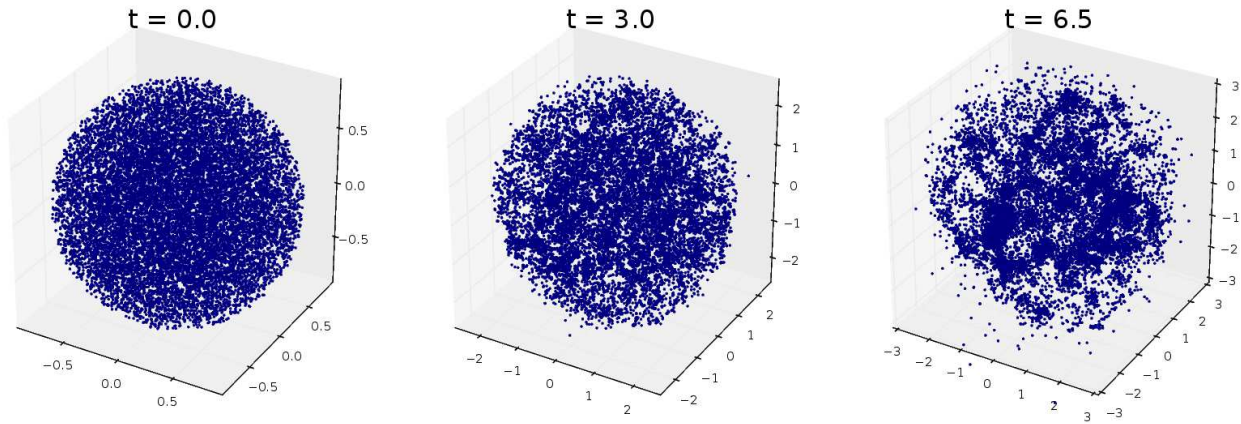}
\caption{Progressive fragmentation through the Hubble expansion. Axis ranges were chosen to preserve the model's aspect. The left panel shows the initial uniform sphere; the middle panel, an intermediate step, slightly fragmented with a much slowed down expansion; the right panel is the final stage, when the expansion has stopped and the fragmentation is fully developed. N=15000 particles were used in this N-body model. Time is in H\'enon units (H.u in the following).}
\label{Fig:fragmentation}
\end{center}
\end{figure*}

\section{Introduction}
\label{Sect:Introduction}
 
The formation of stars in clusters and associations is arguably an important channel for the photometric and chemical evolution of the host galaxy (e.g. \citealt{ladalada03,weidner05,SPZ10}). In the Milky Way, deep IR surveys have long revealed that young stellar associations cover a wide-range in morphology and density (e.g., \citealt{offner14} for a review), as well as probing the low-mass end of the stellar population  (see e.g. \citealt{moraux03,moraux07,andre13}). Examples of complex/irregular morphologies include the Aquila  and $\rho$ Ophiuchus regions \citep{andre07,andre14} as well as the Taurus-Auriga association. Observations by \cite{cortes10}  of the W43 star forming region show a clump of gas in the process of collapse within a highly fragmented broader region (\textit{e.g.} \citealt{motte03}). Denser clusters show less sub-structures (or, clumps) than more extended ones, pointing to  dynamical interactions and mixing on a short timescale to smooth out irregularities: a prime example of this is the Trapezium region of the ONC cluster \citep{hillenbrand98}. Others,such as IC348, may even be currently in this relaxation phase \citep{cambresy06,cottaar15}.

This rough picture of cluster formation and early evolution is backed up to some extent by computer simulations of fragmentation modes in the turbulent ISM \citep{klessen00,maclow04}. In that `bottom-up' picture of cluster formation, turbulent modes decay rapidly once the energy source dies out. Runaway cooling leads to the formation of several fragments which develop along filaments and in knots of high concentration. If the background tidal field is weak, and the star forming region sits well inside its Roche radius, the clumps then converge to the system barycentre  and form a unique,  relaxed  self-bound association over a  course of a few crossing time (see e.g. \citealt{allison09,parker11,maschberger10,bate14,fujii12}~;  see also the discussions in \citealt{andre14,offner14}). 

Several theoretical studies have set out to quantify the characteristics in equilibrium of such associations and clusters for comparison with observations \citep{boily99,goodwin04,mcmillan07,allison09,vesperini14,caputo14} \footnote{Note that similar studies carried out as far back as the 1980s were concerned by the collision-less evolution of such systems, in an application to the formation of elliptical galaxies \citep{vanalbada82,mcglynn84}, which was possible at the time using low-resolution numerical models.}. 
In more recent studies, high-precision orbit integration powered by GPU-accelerated platforms  allows a statistical sampling of star-by-star calculations from a few tens and up to $\sim 10^5$ member stars \citep{allison09,becker13,vesperini14,caputo14}. One of the main emphasis of these sudies is that mass segregation between the stars operates on a short, dynamical timescale, i.e. during the very formation of the associations \citep{allison09,caputo14}. Gravitational collisions between proto-stars should shape up the distribution function of stars  (\citealt{reipurth01,marks12,becker13}~; see \citealt{reipurth14} for a review). \cite{maschberger11} and \cite{moeckel11} have noted that massive stars tend to sit at the heart of gas clumps in hydrodynamical simulations, some as the result of merger events with low-mass proto-stars.       Recently, \cite{fujii16} and \cite{moeckel12} developed hybrid methods using the outcome of hydrodynamical simulations to spawn initial conditions for further stellar dynamical evolution.

The difficulty  to bridge over self-consistently from the star formation phase, to the equilibrium configuration of bound clusters, remains a major hurdle~: hydrodynamical calculations of star forming regions evolve for  up to a few $\times 10^5$ years, when a stable configuration would require several $\times 10^6$ years at typical cluster densities of $10^4$ to $10^5$ stars per cubic parsec. A way to overcome this issue is to run simulations of the dynamical evolution of stellar clusters starting with initial conditions that correspond to the outcome of hydrodynamical calculations. In that context the method of \cite{goodwin04} to setup fractal-like cluster configurations has proved very fruitful in shedding light, for instance, on the development of mass segregation between stars \citep{allison09,parker11,parker14}. \cite{kupper11} extended this method to specific radial density profiles. 
  One drawback from these and similar methods is that the velocity field is not (fully) self-consistent with the inner structure of individual stellar subgroups, at least when done by randomly 
sampling a chosen velocity distribution function. Instead one would expect that the stellar orbits  be self-consistent with the local mass distribution on a scale where several dynamical times match the formation time of individual stars. Also, the spatial coordinates should correlate with mass \citep{moeckel11}, an aspect not readily included in the spherically-symmetric computer models of, e.g. \cite{caputo14} for the dense and young LMC cluster R136. 
This brings up questions about the early evolution of  R136-like clusters, because a cluster forming from the assembly of mass-segregated clumps would lead to a top-heavy stellar MF at birth and alter the expected subsequent evolution of the cluster (mass profile, mass loss, photometric colours) compared with one starting from uniformly distributed stars \citep[Fig 3 \& 5]{haghi14}.

To take into account the stellar interactions occurring during the cluster formation process, we implemented a method whereby stellar density fluctuations are allowed to develop and form self-bound clumps of stars. We start from an uniform sphere in which we introduce a radial velocity field, akin to the Hubble flow in cosmology\footnote{We keep to the conventional name and syntax but take stock of the key contribution of G. Lema\^itre in  the discovery of the cosmological expansion (see \citealt{freeman15}.)}. The growth of density fluctuations follows from the adiabatic gravitational cooling driven by the expansion. While this is not a substitute for fully consistent hydrodynamical simulations, this will lead to a complex configuration with stellar substructures corresponding to filaments and knots and, crucially, a self-consistent velocity field. If we adopt the configuration at the end of the adiabatic expansion as initial conditions for subsequent dynamical evolution, we can then move forward in time and study the evolution to equilibrium of that configuration.

The layout of the paper is as follows. \S\ref{sect:Hubble} revisits the problem of the growth of density fluctuations in an expanding system. This helps to understand and quantify the level of substructures at the end of expansion in relation with the internal structure of the stellar clumps. \S\ref{Sec:Timescales} provides estimations of useful timescales related to the system. A quantitative example of the growth of density fluctuations during the Hubble expansion phase is given in \S \ref{Sec:HubbleExpansion}. We build a range of clumpy models and present an analysis of their properties in terms of mass segregation, mass spectrum and internal velocity in \S\ref{Sec:Nbody}.

The fragmented models built and analyzed in the previous sections are then used as initial conditions in N-body computer simulations to understand their dynamical evolution until equilibrium; the results are analysed in \S\ref{Sec:collapse} and \S\ref{Sec:Segregation}. Finally we discuss the results and look out to future work in \S\ref{Sec:discussion}. All N-body calculations were made using the collisional stellar dynamical code Nbody6 \citep{aarseth03}.


\section{Initial conditions for fragmented stellar systems}
\label{sect:Hubble}
To build coherent highly fragmented stellar systems, we took a hint from \cite{klessen00} who applied periodic boundary conditions to solve the hydrodynamical equations in the Zel'dovich approximation. Their idea was to speed up their SPH calculations by first matching the density and velocity field to first order in density fluctuations. It occurred to us that one needs not to stop at first order, and may instead allow for the full development of density fluctuations under gravity only. Thus we treat mass elements as stars and allow Poissonian density fluctuations to grow until small individual stellar clumps reach equilibrium. Figure \ref{Fig:fragmentation} illustrates the basics of the fragmentation process for an N=15000 stars model. Our view is that by following through with a full stellar IMF, the massive stars will define a radius of influence around themselves and sit preferentially, but not systematically, at the heart of sub-structures, retrieving a feature seen in star formation calculations \citep{moeckel11,maschberger10}, while short-cutting  costly computer calculations. The configuration that is sought here is not unlike the situation found in the formation of galaxy clusters in cosmology, as massive galaxies tend to drag in less massive ones and sit at the heart of clusters.

\subsection{Hubble flow}
\newcommand{\Mtot}{{\cal M}}
We setup a uniform sphere of mass $\Mtot$, bounding radius $R_o$ with $N$ mass elements, corresponding to stars, drawn from a Salpeter mass function and uniformly in space. We give each element an outward radial velocity so $\bold{v} = \Hub_o \bold{r}$, with $\Hub_o$ a ``Hubble-like" parameter chosen such that the total system energy E ($W$ is the potential energy and $E_k$ the kinetic energy) reads:

\begin{table}
\begin{center}
\caption{Summary of main variables.}
\label{Tab:identities}
\begin{tabularx}{\columnwidth}{rl}
\hline
$E$ & Total system energy \\
$E_*$ & Dimensionless total energy \\
$W$ & Total potential energy \\
$E_k$ & Total kinetic energy\\
$\cal M$ & Total system mass\\
$R_o$ & Initial bouding radius\\
$\Hub_o$ & Initial Hubble parameter\\
$v_o$ & Initial velocity at bounding radius\\
$\Hub$ & Variable Hubble parameter\\
$\tau$ & Dimensionless time\\
$x$ & Comoving spatial coordinate\\
$a(t)$ & Rescaling function\\
$\theta$ & Calculation angle\\
$\nu(\tau)$ & Dimensionless velocity $1 + 2E_*(1/a(\tau)-1)$\\
$\xi$ & Radial displacement from comoving \\
$\delta\rho,\delta M,\delta\rho$ & Perturbed quantities\\
$\mu(\tau)$ & Central point mass\\
$\eta$ & Peculiar velocity $d\xi/dt$\\
\hline
\end{tabularx}
\end{center}
\end{table}

\begin{equation}\label{Eqn:HubbleEnergy}
 E = W + E_k = - \frac{3}{5}\frac{G\Mtot^2}{R_o} + \frac{1}{2} \Hub_o^2 \frac{3\, \Mtot R_o^2}{5} \le 0 .
\end{equation}
In the mean-field approximation, the mass inside any shell of radius $r(t)$ is conserved as they move outwards. 
The position of a mass element is known in parametric form from a rescaling of its initial coordinates and we may write 

\begin{equation} \bold{r}(t) = a(t)\bold{x}\, ;\,\bold{v}(t) = \dot{a}\bold{x} = \Hub(t)\bold{r}\ , \end{equation} 
 where $\bold{x}$ is a co-moving coordinate of position, and $a(t)$ is a dimensionless function of time. The flow is homological and no shell-crossing takes place. It is convenient to introduce a dimensionless time $\tau$ such that 
 
 \begin{equation} \label{Eqn:taudef}
  t = \tau / \Hub_o .
 \end{equation} 
 
  We then have
\begin{equation}\label{Eqn:Expansion} 
\left. \left[ \frac{2 E_\ast}{2E_\ast -1} \right]^{\frac{3}{2}} \, \left[ 2\theta - \sin{2\theta} \right]\, \right\vert_{\theta_o}^\theta =  2\sqrt{2E_\ast} \tau 
\end{equation}
where we have defined 
\begin{subequations}\label{Eqn:Scales} 
	\begin{equation}  \label{Eqn:Scalesa} 
		a(t)  \equiv  \frac{\sin^2\theta(\tau)} {\sin^2\theta_o}  
	\end{equation}
	\begin{equation}  \label{Eqn:Scalesb} 
		v_o \equiv \Hub_o \, R_o 
	\end{equation} 
	\begin{equation}  \label{Eqn:Scalesc} 
		E_\ast \equiv \frac{G\Mtot}{v_o^2 R_o } \ . 
	\end{equation}
\end{subequations}

The dimensionless energy parameter $E_\ast$ satisfies  $2 E_\ast  > 1$ for bound systems. The origin of time $\tau = 0 $ coincides with  the angle $\theta_o$ found from solving 
 $\sin^2\theta_o = (2E_\ast - 1) / 2 E_\ast $. 
The solution (\ref{Eqn:Expansion})  
provides the time-sequence for the position and velocity of any shell $ 0 < x < R_o$ as parametric functions of $\tau$~: 

\begin{subequations}\label{Eqn:Hubble} 
\begin{equation}
 v^2(t) = \Hub_o^2 x^2 \, \left( 1 + 2E_\ast [ 1/a(\tau) - 1 ] \right) \equiv \Hub_o^2 x^2 \nu^2(\tau)
 \end{equation} 
 \begin{equation} 
  \Hub(t) = \Hub_o \, \nu(\tau) / a(\tau) 
 \end{equation} 
\begin{equation}
\rho(t) = \frac{3\Mtot}{4\pi R_o^3}\frac{1}{a^3(\tau)} \equiv \rho_o \, n(\tau) \ . 
\end{equation}			
\end{subequations}

The derivation of the expansion end-time is detailed in Appendix \ref{App:Hubble}.
 
\subsection{Fragmentation modes}
\label{Sec:FragmentationModes}
\newcommand{\xistar}{\xi_\ast}
\newcommand{\xstar}{x_\ast}
\newcommand{\etastar}{\eta_\ast}
\newcommand{\Estar}{E_\ast} 
It is instructive to follow what happens to density perturbation modes in the expanding uniform sphere described by equations (\ref{Eqn:Expansion}) and (\ref{Eqn:Scales}), as the local density increase also gauges the rise in velocity dispersion. A simplified 
calculation for radial modes of perturbation in the linear approximation will be derived here, with the goal to determine when the clumps become mostly self-gravitating. A more detailed analysis can be found in the classic work by \cite{friedman78}, \cite{peebles80} and \cite{aarseth88} .

We introduce a Lagrangian perturbation in the position of a shell of constant mass by substituting $\mathbf{x} \rightarrow \mathbf{x} + \mathbf{\xi}(\mathbf{x},t)$ and we set $\mathbf{\xi} = \xi \hat{\mathbf{r}}$ for a radial displacement. A linear treatment of the continuity equation yields the perturbed density 

\newcommand{\delrho}{\delta\rho} 
\begin{equation} \label{Eqn:Deltarho} 
\delrho = - \mathbf{\nabla}\cdot (a\rho\mathbf{\xi}) =  - \rho(\tau) \frac{1}{x^2}\frac{\partial}{\partial x} ( x^2\xi ) 
\end{equation}
which leads to a perturbation in  the mass integrated up to  radius $r$ 
\[ \delta M(<r) = - 4\pi a^3(\tau) \rho x^2 \xi \ . \] Poisson's equation in spherical symmetry gives the perturbed potential 

\newcommand{\delphi}{\delta\phi} 
\begin{equation} \label{Eqn:Poisson} 
\frac{1}{r^2}\frac{\partial}{\partial r} r^2\frac{\partial}{\partial r} \delphi =\frac{1}{a^2}\frac{1}{x^2}\frac{\partial}{\partial x} x^2\frac{\partial}{\partial x} \delphi  = 4\pi G\delrho .
\end{equation}
Subtituting for $\delrho$ from (\ref{Eqn:Deltarho}) in (\ref{Eqn:Poisson}) and integrating once, we obtain the general solution

\begin{equation}\label{Eqn:Gradpsi} 
a(\tau)\nabla \delphi = \frac{3G\Mtot}{R_o^3} \left( - \xi + R_o^3\frac{\mu(\tau)}{x^2} \right) 
\end{equation}
where $\mu$ stands for a central point mass. A point mass would form by shell crossing at the center of coordinates. In an expanding system, shell crossing at the center is unlikely. For that reason, we make $ \mu = 0$ in the remainder of this paper.

The equations of motion at co-moving radius $x +\xi(x,t)$ can be expanded to first order in $\xi$~; identifying terms of the same order we obtain (with $\partial/\partial x = \nabla_x$)

\begin{equation} 
a(\tau) \frac{d^2}{d t^2} \xi + 2 \dot{a}(\tau) \frac{d}{dt}\xi = - \nabla\delphi - \xi \nabla_x \nabla\phi - \ddot{a}(\tau)\xi \ . 
\end{equation} 
The second and third terms on the right-hand side cancel out exactly~; the first is known from (\ref{Eqn:Gradpsi}). 
It is standard practice to demote this second-order dynamical equation to a set of first order equations~; 
for convenience we use the initial system radius $R_o$ as unit of length,  and we introduce starred ($\ast$) dimensionless variables. We then have $x = R_o x_\ast, \xi = R_o\xistar$, and so on.  After simplification using the dimensionless functions of $\tau$  defined in (\ref{Eqn:taudef}), the differential equations read

\begin{subequations}
 \label{Eqn:Equations}
    \begin{align}
		\frac{d}{d\tau} \xistar &=  \etastar(\tau)  \\ 
		\frac{d}{d\tau} \etastar &= \frac{3 \Estar}{a(\tau)^2}\,\xistar - 2 		\frac{\nu(\tau)}{a(\tau)} \etastar   
	\end{align}
\end{subequations}
where we have introduced the peculiar velocity $\eta \equiv d\xi/dt = \Hub_oR_o \eta_\ast$.  

\newcommand{\xistaro}{\xistar^{(o)}}

\subsection{Integration} 
\subsubsection{Initial conditions} 
Equations (\ref{Eqn:Equations}) 
are easily integrated with an explicit integration scheme once the initial values $R_o, \Hub_o, {\cal M}$ and $\xistar(0)$ are specified~;  all 
functions of the dimensionless time $\tau$ are set to unity except that $\etastar(0) = 0$. 
The Hubble parameter $\Hub(\tau) \rightarrow 0$ when the system reaches a 
maximum radius $a(\tau)R_o$ ($\theta[\tau] = \pi/2$ in Eq.~\ref{Eqn:Scales}).
 Around that time, the Lagrangian displacement $\xistar$ grows exponentially, and the clumps become the densest (see Eq. [\ref{Eqn:Equations}]).  
 We investigate the growth of a density perturbation as a Fourier  fragmentation mode. In the linear regime, such a mode is decoupled from all the others. We pick 

\begin{equation} \label{Eqn:Fouriermode} 
   \xistar(x,0) = \xistaro \sin( kx ) ,
\end{equation}  
where the wavenumber $k$ is such that $k R_o = m \pi$ and $\xistar(R_o,0) = \xistar(R_o,\tau) = 0$ at all times. When deciding 
which wavenumber to choose, we must bear in mind the finite numerical resolution of the models that we will present later. The next subsection 
gives quantitative arguments that motivated our choices.  

\subsubsection{Fourier modes: resolution issues} 
\label{Sec:FourierModes}
For a uniform distribution of $N$ discrete mass elements, the 
mean separation $l_o\simeq R_o / N^{1/3}$ gives a reference wavelength $\lambda / R_o = \lambda_\ast \ge N^{-1/3}$ for a resolved  Fourier mode. 
Since $kR_o = m\pi$, this also implies that $ m \le 2 N^{1/3}$.  Poisson statistics 
help set the initial amplitude $\xistaro$  of a perturbation. The radius bounding a shell of $N$ mass elements distributed randomly will fluctuate 
freely between $r$, $r + \delta r$ due to stochasticity. The radius $r$ of a uniform sphere being a power-law of mass $M$, we 
find $\delta r/r = 1/3\, \delta M/M = 1/3\, \delta N/N = 1/3\, N^{-1/2}$ for identical mass elements. 
We then compute the number-averaged value $\langle \delta r/r\rangle$ by summing over  all elements from 1 to $N$ and dividing by $N-1$ to find 

\[ \langle \frac{\delta r}{r} \rangle = \langle \xistaro\rangle = \frac{2}{3} \frac{\sqrt{N} - 1 }{N - 1} \ . \]
Thus the mean amplitude (in units of $R_o$) is $\langle\xistaro\rangle\simeq 1/10$ for $N=32$ which drops to $\langle\xistaro\rangle\simeq 6\times 10^{-4}$ when $N = 10^6$. We checked that the mode with the shortest wavelength $\lambda_\ast$ still resolved would have a displacement $\langle\xistaro\rangle$ initially 
smaller than $\lambda_\ast/2$ for any sensible value of $N$. This in turn implies that this mode may grow over time to reach an amplitude $\xistar(x,\tau) \simeq \lambda_\ast/2$, which is the point when orbit-crossing between shells of constant mass must occur. If we isolate for the time when shell-crossing occurs, we can then explore whether the clump just formed (the high-density, non-linear 
dynamics of which is not covered by Eqs.~[\ref{Eqn:Equations}]) can exhibit two-body relaxation effects, due to its internal dynamics. To answer that question,  
below we introduce reference timescales for two-body relaxation in stellar systems.  

\section{Timescales} 
\label{Sec:Timescales} 
We already noted that $\Hub_o^{-1}$ sets a time-scale for the expansion of the system. That time should be chosen so that it matches the hydrodynamical star formation
phase of $0.5 - 1$ Myr \citep{maschberger11,bate14}. 
When $\Hub(\tau) = 0$ and the expansion is over, the stars relax to a new equilibrium driven by star-star interactions. Therefore we need to address first the 
internal dynamics in clumps in time units of $\Hub_o^{-1}$, before discussing the later phase of violent relaxation and consider the system as a whole. 
The definitions are the same, only the face values change between the two phases of evolution. 

\newcommand{\trel}{t_{rel}}
\newcommand{\tc}{t_{cr}} 
\newcommand{\tms}{t_{ms}} 

\subsection{Relaxation- and dynamical timescales} 
We consider a clump of stars drawn from a mass spectrum, linked together by self-gravity, and  follow standard definitions \citep{meylan97,fleck06} for the system crossing time $\tc$ as 

\begin{equation}\label{Eqn:Tcross} 
     \tc = \frac{2r_h}{\sigma} = \frac{2 r_h}{\sqrt{GM_\lambda/r_g} } \, ,
\end{equation}
where $r_h$ is the half-mass radius, $\sigma$ the three-dimensional velocity dispersion, $M_\lambda$ the mass of the clump of gravitational radius $r_g$ 
given by $GM_\lambda / r_g = \sigma^2$. We call the two-body relaxation time $\trel$ the ratio 

\begin{equation} \label{Eqn:Trel}
\frac{0.138}{2} \left(\frac{r_h}{r_g}\right)^{1/2} \frac{N_\lambda}{\ln 0.4 N_\lambda}  \equiv  \frac{\trel}{\tc} .
\end{equation}
We set $M_\lambda = N_\lambda \langle m_\star\rangle $ where the mean stellar mass $\langle m_\star\rangle$ will be computed from the stellar IMF (see \S\ref{Sec:Scales} below). 

\subsection{Mass segregation time-scale}
The spread of stellar masses and the trend towards equipartition of kinetic energy will enhance evolution in the orbits of the stars as heavy stars 
segregate to the barycentre of the system \citep{meylan97}. 
We borrow the definition of \cite{fleck06} for the mass-segregation timescale and write

\begin{equation}\label{Eqn:Tms} 
  \frac{\tms}{\trel} \equiv \frac{\pi}{3} \frac{\langle m_\star\rangle}{\max\{m_\star\}} \, \frac{\bar{\rho}_h}{\rho_g} \left( \frac{r_h}{r_g}\right)^{3/2} ,
\end{equation}
where $\bar{\rho}_h = M_\lambda/2 /( 4\pi/3) r_h^3$ is the mean density within radius $r_h$, and $\rho_g$ the mean density inside a sphere of 
radius $r_g$. For a clump of total density $\rho + \delrho$ given by Eq.~(\ref{Eqn:Deltarho}), we may write 

\begin{equation}
 \rho_g = \frac{\rho_o}{a^3(\tau)} \, \left( 1 + \frac{\delrho}{\rho} \right) \equiv  \frac{\rho_o}{a^3(\tau)} \, \rho_\ast.
 \end{equation}
Combining this with Eqs.~(\ref{Eqn:Tcross}), (\ref{Eqn:Trel}) and (\ref{Eqn:Tms}), the mass-segregation timescale now reads 

\begin{equation}
 \tms = \frac{0.138}{6}\pi \left(\frac{3}{4\pi}\right)^{1/2} \frac{\langle m_\star\rangle}{\max\{m_\star\}} \, \frac{N_\lambda}{\ln 0.4 N_\lambda} \, (G\rho_g)^{-\frac{1}{2}}\, . \end{equation}
  
Making use of the equality 

\begin{equation}
\frac{4\pi}{3} G\rho_o = \Hub_o^2 \Estar ,
\end{equation} 
the last three relations simplify to 

\begin{equation}\label{Eqn:Taums} 
\Hub_o \tms = \frac{0.138}{6} \pi \, \frac{a_\lambda^{3/2}}{(\rho_\ast\Estar)^{1/2}} \, \frac{\langle m_\star\rangle}{\max\{m_\star\}} \, \frac{N_\lambda}{\ln 0.4 N_\lambda} \equiv \delta\tau_{ms}
\end{equation}
where $a_\lambda$ refers to the expansion factor $a(\tau)$ evaluated at time $\tau$ when $\xistar \simeq \lambda_\ast/2$ and $\delta\tau_{ms}$ is the dimensionless segregation time-scale. Note that our use of Eq.~(\ref{Eqn:Deltarho}) to compute $\rho_g$ means that the gravitational radius $r_g$ does not have its usual definition based on the gravitational energy $W$ of the system. Linking 
$\rho_g$ to $r_g$ in this way has the advantage that $r_g$ is not derived from an implied mass profile, which is (by definition) not resolved 
here. 

Clearly the segregation time depends strongly on the mass spectrum of individual clumps, on their membership $N_\lambda$, as well as the density contrast $\rho_\ast(\tau_\lambda)$. We find the density contrast from (\ref{Eqn:Fouriermode}) and (\ref{Eqn:Deltarho}),  

\[ \left.\frac{\delrho}{\rho}\right|_{\tau=0} \!\!\!\! = - \frac{1}{x^2}\frac{\partial}{\partial x^2} x^2\xi = - \left( 2 \frac{\sin\, m\pi x_\ast }{m\pi x_\ast} + \cos\,m\pi x_\ast \right) m\pi   \xistaro
\]
which admits an upper-bound of $3 m\pi \xistaro$. In the course of evolution, the initial amplitude of perturbation grows to $\xistar = \lambda_\ast/2$ so that the density peaks at 

\begin{equation} \label{Eqn:Densitypeak} 
  \rho_\ast = 1 + \frac{\delrho}{\rho} = 1 + 3 m\pi \lambda_\ast / 2 = 1 + 3\pi\, ,
\end{equation}
where the last substitution follows from the definition of the integer $m$. 
The mass $M_\lambda$ in a shell bounded by $r, r+ \lambda$, is known from the unperturbed density profile~; in terms of the total system mass ${\cal M}$, we find 

\begin{equation} \label{Eqn:Clumpmass} 
   \frac{M_\lambda}{{\cal M} } = ( \overline{3 x_\ast^2} + \lambda_\ast^2/4 ) \lambda_\ast = ( 1 + \lambda_\ast^2/4) \lambda_\ast \, , 
\end{equation}
where we have replaced $3x_\ast^2$ by its space-averaged value in  the last step.  Eq. (\ref{Eqn:Clumpmass}) provides an estimate of  bound mass of a clump formed through the growth of a radial perturbation mode. If all the stars have equal mass, or, if the stellar mass 
function is symmetric with respect to the mean value $\langle m_\ast\rangle$, the ratio of the number $N_\lambda$ of stars in the clump to the total number $N$ is in the same 
proportion as $\frac{M_\lambda}{\cal M}$. We argued in \S2.3.2 that a resolved mode should have $\lambda_\ast \ge N^{-1/3}$. Putting this together we find an estimate for $N_\lambda$ 
which reads 

\begin{equation} \label{Eqn:Clumpn} 
   N_\lambda = N^{2/3}\left( 1 + \frac{N^{-2/3}}{4} \right) .
\end{equation}
This number inserted into Eq.(\ref{Eqn:Taums}) leads to a rough picture of the segregation process in clumps. The rate of mass segregation leans  on the choice of initial value for the expansion phase, $\Hub_o$. In the limit when $\Hub_o = 0$, there is no expansion whatsoever, and the clumps form unsegregated (aside from random associations when attributing positions and velocities to the stars) during global infall. If by contrast, the expansion is vigourous, $a_\lambda \gg 1$, and the segregation timescale remains large. For $N \sim 10^4$, we compute from (\ref{Eqn:Clumpn}) $N_\lambda \approx 464$: a clump with that many stars will mass-segregate rapidly only if its stellar mass function includes very massive stars. We note that one-dimensional (radial) modes would in fact split into several smaller fragments in a three-dimensional calculation.\footnote{ A full-grown radial mode forms a thin shell subject to fragmentation. See \textit{e.g.}  \cite{ehlerova97,wunsch10}.} We expect  the clumps to form quickly  and contain $N_\lambda \ll 464$ stars, so  that the internal dynamics 
will drive mass segregation {\it before} the system expansion stops. Because this depends in the details on $\Hub_o$ and other important parameters, 
we defer the analysis to \S \ref{Sec:Nbody} and N-body simulations. 


\section{Fragmentation  during the Hubble expansion phase} 
\label{Sec:HubbleExpansion}
This section discusses the growth of fragments and their properties at the end of the Hubble expansion phase. Because hydrodynamical calculations of supersonic turbulence show proto-stars forming during a single free-fall time,  

\begin{equation} \label{Eqn:Freefall} 
   t_{ff} = \left(\frac{3\pi}{32 G\rho_o} \right)^{1/2} \sim 0.5 - 1 {\rm Myr}
\end{equation}
(where $\rho_o$ is taken as the global mean density), we should pick a set of parameters such that the clumps form over a physical time to match that of Eq.~(\ref{Eqn:Freefall}). In this paper, we choose to evolve the models until $H = 0$ to allow for fully-developed individual clumps. More precisely, we look for a computational setup such that $\Hub(\tau) = 0$ in a minimum of one star-formation timescale $t_{ff}$. We do so although the approach taken here to form clumps would allow to stop the calculation {\it before} $H = 0 $ and {\it subtract} the residual radial velocities using Eqs \ref{Eqn:Hubble}. As a result, the configuration would be less fragmented than for the case when $H =0$. Thereafter the dynamical evolution would proceed similarly in all the cases, but with different clump mass- and size distributions. The choice of initial Hubble parameter $\Hub_o$ must always yield $E < 0$ in (1). Note again that when $\Hub_o$ is set equal to zero, we recover the classic configuration for the cold collapse of uniform bodies.

\subsection{Choice of scales, mass function} 
\label{Sec:Scales} 
To ease comparisons with N-body calculations cast in standard \cite{henon73} units (see also \citealt{heggie86}), we set ${\cal M} = G = R_o = 1 $ and use $\Hub_o = 1.0833 .. \simeq 1$ so that the total binding energy $E = -1/4$. The Hubble expansion proceeds until a time $t = \tau / \Hub_o \simeq 3.87 / \Hub_o$, when $H = 0$ and the bounding radius $R$ reaches $R = a(\tau)R_o \simeq 2.4 R_o$. The evolution time up to that point coincides almost exactly with the {\it current} global system free-fall time of $\approx 4.1$ time units. System-wide collapse to the barycentre will ensue on the same time-scale, but now this process will involve the merging / scattering of several high-density clumps. 

The mass of individual stars follow a  truncated \cite{salpeter55} distribution function, where the distribution function $ dN/dm \propto m_\star^{-\alpha}$ with index $\alpha = 2.35$ for masses in the range  $0.3 M_\odot < m_\star < 100 M_\odot$ giving a mean value of $\simeq 1 M_\odot$. We chose this form mainly for simplicity, and for ease of calculations. 
A more realistic stellar mass function should be used in future work \citep{chabrier05,kroupa02} and include multiple stars \citep{marks12} for more quantitative  comparisons with observations.

%
\begin{figure}
\begin{center}
\includegraphics[width=\columnwidth]{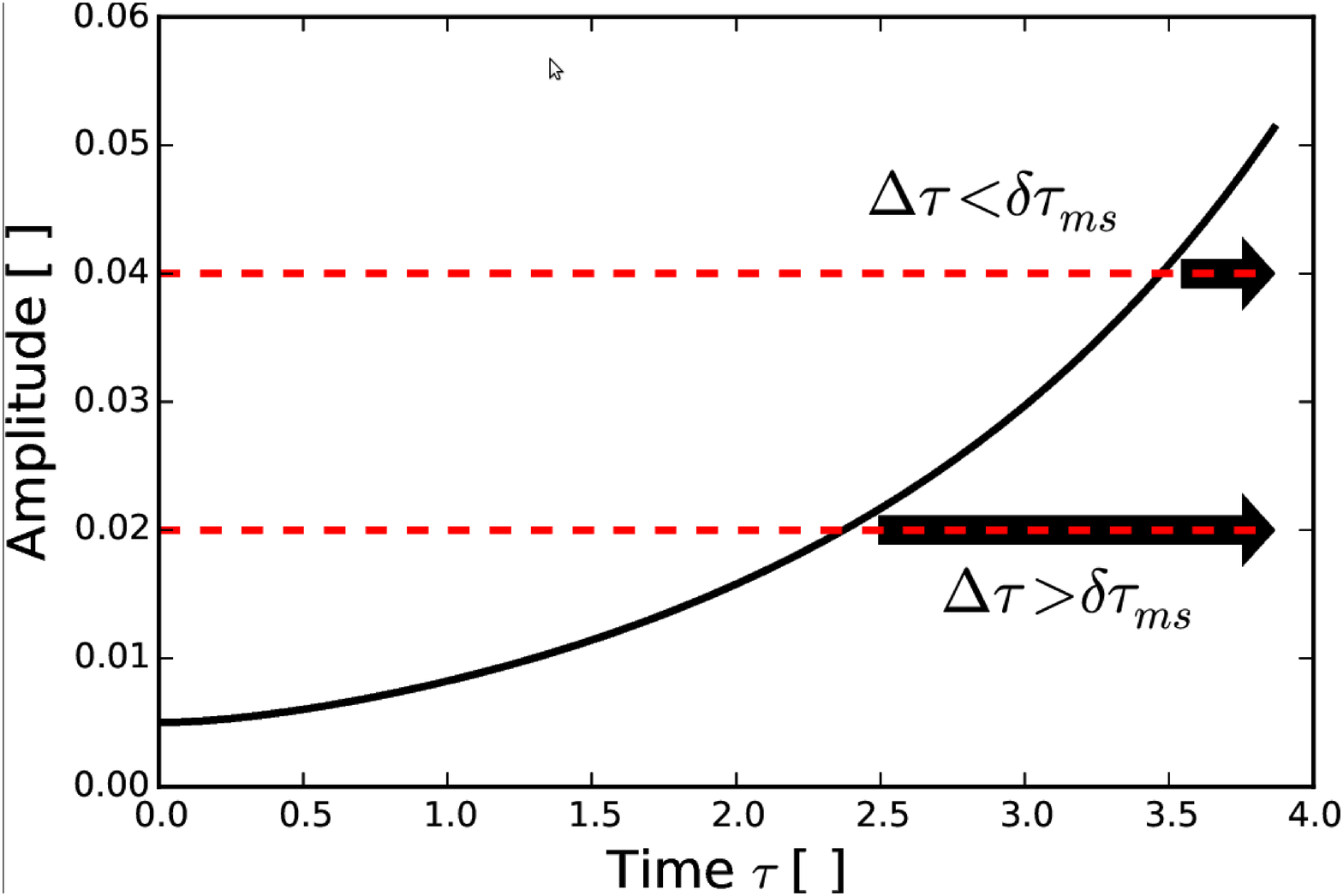}
\caption{Time-evolution of the perturbation amplitude $\xistar$ for a system with $N = 15 000$ stars. The radial expansion of the system ends at $\tau = 3.86$\,. The amplitude $\xistar$ matches the shortest wavelength $\lambda_\ast/2 \simeq 0.02$ for $\tau \simeq 2.3$ (see the horizontal dashed lines). The remaining time-interval $\Delta\tau \simeq 3.86 - 2.3 = 1.56 \gg \delta\tau_{ms}$ estimated from (\ref{Eqn:Taums}). The same is not true for a mode of wavelength such that $\lambda_\ast/2 = 0.04$~: modes of large wavelength tend to produce less mass-segregated clumps.}
\label{Fig:Amplitude}
\end{center}
\end{figure}

\subsection{Example with N = 15000 stars}
\label{Sec:Example}
We describe the evolution of a system through the perturbation equations before turning to N-body modelling in \S\ref{Sec:Nbody} . Let us fix $\Estar = 6/7$, with $\Hub_o = 1$ 
and set $N = 15 000$ as reference\footnote{The more accurate value is $\Hub_o = 1 + 1/12 = 1.0833$ but we rounded up to 1 to simplify the discussion}. We compute a mean initial amplitude of perturbation $\xistaro \approx 0.005 $ with a shortest-resolved wavelength $\lambda_\ast \approx 0.04$. Fig.~\ref{Fig:Amplitude} displays the solution from integrating Eqs.~(\ref{Eqn:Equations}). The amplitude $\xistar(\tau)$ grows monotonically and crosses the values $\lambda_\ast/2$ at $\tau \approx 2.3$~: thereafter the perturbation enters a non-linear regime of evolution during which the internal dynamics may become collisional $( \Delta\tau > \delta\tau_{ms})$. A second case is depicted on Fig.~\ref{Fig:Amplitude}, where the wavelength $\lambda_\ast = 0.08$ and the perturbation reaches amplitude $\xistar = \lambda_\ast/2$ at $\tau \approx 3.6$: there is then too little time left before the end of the Hubble expansion phase for a clump of stars to evolve collisionally ($\Delta\tau < \delta\tau_{ms}$). 

The dynamical state of individual clumps is clearly a question of membership $N_\lambda$ and mass spectrum as shown in (\ref{Eqn:Taums}). We have been arguing that most small-size clumps will show collisional internal evolution~: a small cluster of stars would lose low-mass stars in the process and so have an increased ratio of average-    to maximum stellar mass. It is not clear, then, whether this trend is strong enough to compensate for the (almost) linear dependance on membership. Anticipating the results of the next section, we take $N_\lambda$ from Eq.~(\ref{Eqn:Clumpn}) to compute  a product  $N_\lambda / \ln 0.4 N_\lambda \times \langle m_\ast\rangle / \max(m_\ast) \sim 14 $ for the case of a Salpeter mass function truncated at $20 M_\odot$ ;  and about $3$ for a truncation value of $100 M_\odot$. In practice, the results of N-body 
calculations yielded values scattered in the range [3, 14] $M_\odot$, 
consistent with there being {\it no trend}  with clump membership $N_\lambda$. To inspect further the actual properties of clumps, we next turn to  N-body calculations.

\section{N-body calculations and analysis} 
\label{Sec:Nbody} 
The stellar dynamics code Nbody6 \citep{aarseth03} treats the gravitational forces of stars with no softening of the potential. The code was ported to GPU platforms \citep{nitadori12} for an effective range of from $\sim 40$ to $\approx 130 000$ stars on a single host computer. The units of computations are those defined in \S\ref{Sec:Scales}. In terms of initial conditions, our approach is similar to what is done in cosmology, with the important distinction that integration is performed in real space, and the evolution of the scale factor $a(\tau)$ is governed by the system's self-gravity (as opposed to being a plug-in). 

We draw $N$ stars from an Salpeter distribution function which we truncate by default to $100 M_\odot$~; in some calculations we will  use a lower bound of $20 M_\odot$. The code preserves the total energy and angular momentum to better than one part in $10^4$ for integration over $\sim 100 $ time units. The numerical integration starts with the expansion phase, but we will refer to the time at which $H = 0$ as $t=0$, as we consider this dynamical state as initial conditions for cluster evolution, see \S\ref{Sec:collapse}.

\begin{table}
\begin{center}
\caption{Summary of fragmentation models and their characteristics. These simulations started from an uniform sphere and were stopped when the expansion halted, at t=3 H.u . The third column shows the number of independent computations for each model.}
\label{Tab:fragmentationmodels}
\begin{tabularx}{\columnwidth}{XXlXX}
\hline
Name & N & Sampling & Mass range  \\
\hline
Rmh20 & 15000 & 30 & [0.35- 20 ]\\
Rmh100 & 15000 & 30 & [0.3 - 100]\\
Rmh1 & 15000 & 30 & 1.0 \\
R40h20 & 40000 & 1 & [0.35- 20 ] \\
R40h100 & 40000 & 1 & [0.3 - 100] \\
R80h100 & 80000 & 1 & [0.3 - 100] \\
\hline
\end{tabularx}
\end{center}
\end{table}

Table \ref{Tab:fragmentationmodels} summarises the main simulations used in this section to investigate the fragmentation of such systems.

\subsection{Clump detection: Minimum Spanning Tree}
\label{Sec:MST}
The study of substructures requires an efficient clump-identification algorithm (or, {\it halo-finding} in cosmology). By clump we mean here a local overdensity of stars. Several methods are commonly used such as the HOP algorithm \citep{eisenstein98,skory10} which relies on attributing local densities to each particle and separating the clumps through density thresholds. The HOP algorithm is very robust on large cosmological data sets. However, our calculations have comparatively coarse statistics and noisy density fields. This issue, coupled with the  large number of free parameters of the HOP algorithm, makes the method less appealing. Instead we follow \cite{maschberger10} who adapted the minimum spanning tree (MST~; see e.g. \citealt{allison09b,olczak11}) technique to the detection of clumps. A spanning tree is a set of edges connecting a group of  particles but without closed loops~; the MST seeks to minimise the total length of the edges. One may then construct the MST for the whole system, and then delete all edges larger than a chosen cutting length, $d_{cut}$. The sub-sets that are still connected  are labeled as clumps. In practice a minimum sub-set size $N_d$  is also chosen so as to avoid many small-N subgroups~: experience led us to choose  $N_d = 12$ for the minimum number of stars per clump.

\begin{figure}
\includegraphics[width=0.5\textwidth]{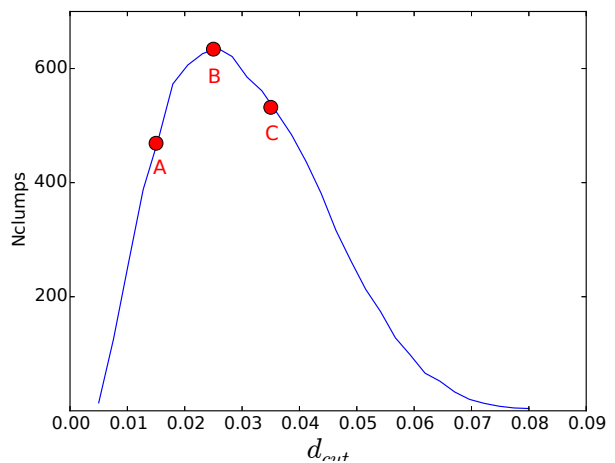}
\caption{Number of clumps detected by the MST algorithm as a function of the cutting length $d_{cut}$.  The data were obtained from model R80h100 (see Table~\ref{Tab:fragmentationmodels}). Points A, B and C are respectively at $d_{cut} = $0.015, 0.025, and 0.035 model units. See also Fig~\ref{Fig:clumpsABC}. $d_{cut}$ is in H.u .  }
\label{Fig:Ndcut}
\end{figure}

\begin{figure}
\begin{center}
\includegraphics[width=0.8\columnwidth]{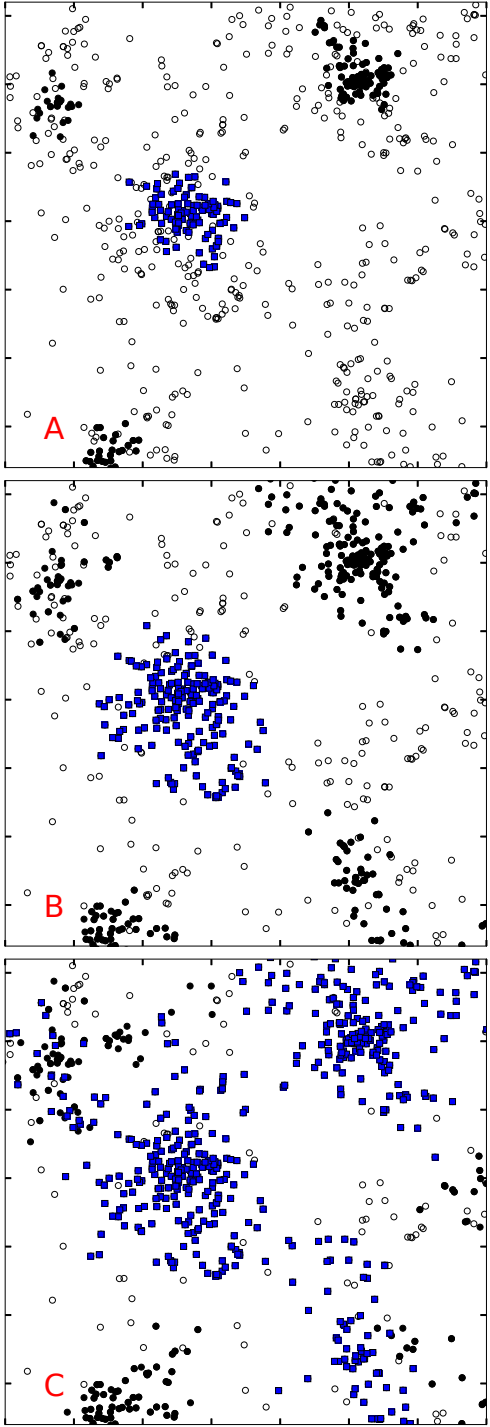}
\end{center}
\caption{Example of detected clumps for three cutting length, 0.015 (top panel), 0.025 (middle panel), 0.035 (bottom panel), which were labeled A,B,C  in Fig.~\ref{Fig:Ndcut}. A cube within the R80h100 model was extracted and projected.  Empty circles are stars which do not belong to any clump, black circles are clump members, and blue squares are stars that are identified as a single large  clump. See text for details. Tick marks are spaced by 0.05 length units for a box size of 0.35 units.}
\label{Fig:clumpsABC}
\end{figure}

With $N_d$ fixed, the length $d_{cut}$ is then the only free parameter left. There is some freedom 
in choosing an appropriate value. 
Maschberger et al. (2010) fixed the value of  $d_{cut}$ by visual inspection of clumps.
 We instead  identified  clumps in a fragmented system for a range of values for $d_{cut}$ and settled for the value  which optimised the number of identifications. This is shown on Fig.~\ref{Fig:Ndcut} for an N = 80k fully-fragmented Hubble model. For small $d_{cut}$'s, the number of detected clumps at first  increases rapidly. The rise is due  to the length $d_{cut}$ initially being small compared with the typical volume spawned by $N_d$ or more  nearest-neighbours. Beyond a certain value, a transition to another regime occurs, whereby the algorithm starts to connect previously separated clumps, counting them as one. The number of clumps thereafter begins to decrease. The value $d_{cut} \approx 0.025$ H.u optimises the outcome of the clump-search. This is a generic feature of the MST algorithm and we have adopted the same strategy throughout, adapting the value of $d_{cut}$ to the number $N$ of stars used. 
   On Fig.~\ref{Fig:clumpsABC}, a sub-set of the model is shown where we have identified stars that belong to clumps with filled symbols. The three panels on that figure are each for a different value of $d_{cut}$, increasing from top to bottom. For the smallest value $d_{cut}$=0.015 H.u, clumps look somewhat truncated as we are still in the under-sampling regime and only their cores registered as clumps. The second, optimal, value $d_{cut}$=0.025 H.u produces visually well-isolated clumps. Finally, the third and  largest value is so that clumps begin to merge together~: this is shown by the unique clump identified in the bottom panel (filled blue squares).

The procedure developed here gives results in agreement with other clump-identification algorithms developed using the MST (see e.g. \citealt{gutermuth09,kirk11}).

\subsection{The velocity field}
\label{sec:velocityfield}

\begin{figure*}
\includegraphics[width=\textwidth]{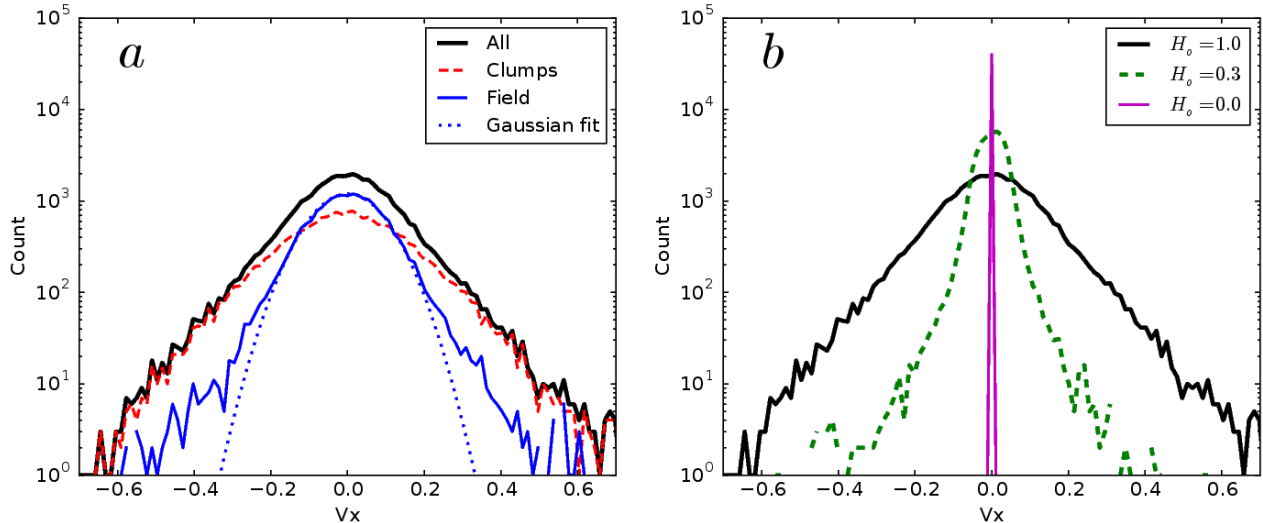}
\caption{a) Distribution of the one-dimensional velocity field for the whole cluster as the thick solid black line, in the simulation labelled as R40h20 in Table~\ref{Tab:fragmentationmodels}, at the time of turnaround ($H = 0$). The red dashed distribution matches clump members and thin solid blue the field particles. b) The distribution for three different values of $\Hub_o$~: when $\Hub_o = 0$, the distribution is a Dirac-$\delta$ around $v = 0$. The central distribution broadens as $\Hub_o$ increases to 0.3 and 1. Observe the exponential profiles at large $|v|$. Velocities are in H.u}
\label{Fig:vxdispersion}
\end{figure*}

One of the aims of this study is to compare the clump configurations derived from the Hubble expansion method with the distribution of proto-stars that form in hydrodynamical simulations. There is no hydrodynamics in the approach that we have taken, nevertheless expansion under gravity alone is equivalent to the adiabatic expansion of gas~: for that case, the first law of thermodynamics equates the drop in internal energy ${\rm d}U$ to minus the external work,  $-p {\rm d}V$. At constant mass, the change in gravitational energy ${\rm d}W$ is $ - {\rm d}E_k$, where $E_k$ is the kinetic energy. With $W < 0 $ but increasing over time, this implies that $E_k$ drops in amplitude. In the case when the motion is strickly radial, $E_k = 0$ when $H = 0$ and all stars come to rest. We ask to what extent the growth of substructures and non-radial motion  off-set the `adiabatic cooling' brought on by expansion. 

Fig.~\ref{Fig:vxdispersion}a graphs the x-axis one-dimensional velocity distribution for a 40k-particle model. The left-hand panel displays the overall distribution as well as the two sub-populations of clumps members and out-of-clump field stars, obtained through the algorithm presented in section \ref{Sec:MST}. We identified some 20944 stars in clumps (or $\approx $ 52\%) at the end of expansion. The expectation that all stars have zero- or low-velocities is validated by the peak in the distribution around $v_x = 0$.

As sub-structures form and  interact mutually, generating tangential as well as radial motion, the peak broadens but remains symmetric about the origin.  The large velocities are brought by stars in clumps, which demonstrates that interactions within the substructures boost the internal velocity dispersion of the cluster as a whole. Field stars dominate the low-amplitude regime. Their velocity distribution is well-fitted with a  Gaussian (shown as a dotted blue line), down to one-tenth the height of the central peak, or about 1\% of all field stars. 
To illustrate further the idea that large velocities are confined to clumps formed by fragmentation modes, we compare on Fig.~\ref{Fig:vxdispersion} a set of models 
with different initial  values of  $\Hub_o$. The right-hand panel on the figure  plots the distribution for three values of $\Hub_o$~: 0, 0.3, and 1. 
Clearly when $\Hub_o=0$, the velocities are identically zero and there is no fragmentation whatsoever (apart from root-N noise). The distribution is then a sharp peak centered on zero. For positive but low values of $\Hub_o$, the fragmentation modes do not develop much before turn-around and the (non-radial) velocities remain small. The central peak  has a much narrower dispersion, and the high-velocity wings are clipped. In this case, too, analysis of the  weakly fragmented system shows that virtually all high-velocity stars are found in clumps. The velocity distribution for the case  $\Hub_o = 1$ is added for comparison. The fact that the full range in velocities is reduced by a factor $\sim 3$ for the 
less fragmented model is also an indication of the shallower potential well of the clumps

The full population velocity distribution (solid black line) at first sight is very similar to those of \citet[Fig.~5]{klessen00}. In that figure, the authors show the velocity distribution of gas particles in a fragmenting system. \citeauthor{klessen00} attribute the high-velocity tails to gas particles falling towards stellar clumps at supersonic speed. Supersonic motions imply that gas particles trace ballistic trajectories, and hence behave like point mass particles. 
A small fraction of field stars in our calculations also have large velocities. 
We suspected that these stars might have acquired their large velocity through in-fall toward a nearby stellar clump. 
We did not, however,  find compelling evidence that would allow us to identify the origin of high velocities in field stars. Inspection of a sequence of snapshots failed to show that the velocity vectors were pointing at nearby stellar clumps: it is therefore not possible to make the same assertion as \citeauthor{klessen00} and state that stellar clumps accrete field stars. 
It is possible, on the other hand,  that high velocities originate from past star-star interactions. However, we did not find clear trends in the few orbits that we studied which would confirm such an event. 
The question of mass accretion by stellar clumps might be best settled if we added gas to our simulations to boost the mass resolution, and analysed model data using mock CCD frames, as did \citeauthor{klessen00}. We defer this analysis to forthcoming work.

We close this section with a remark about the velocity distributions seen on Fig.~\ref{Fig:vxdispersion} and the internal state of the stellar clumps. Because small clumps would have time to evolve dynamically through star-star collisions and reach a state of near-equilibrium (see \S\ref{Sec:Timescales}) we would expect clumps to develop a velocity field similar to Mitchie-King models of  relaxed self-gravitating star clusters \citep{BT08}.  The one-dimensional velocity distribution of Mitchie-King models plotted in a logarithmic scale approaches a flat-top when $|v_{1d}|$ is small,  and cuts-off rapidly at large values~: the distributions are  concave at all velocities. This holds true for all models independently of their King parameter\footnote{Notice how this holds only because of the choice of a logarithmic vertical axis.} $\Psi(0)/\sigma^2$. The shape of the distributions displayed on Fig.~\ref{Fig:vxdispersion}, on the other hand, is convex as we shift, from small, to large $|v_{1d}|$. We deduce from this straightforward observation that the clumps that formed through fragmentation and subsequent mergers cannot be treated as fully in isolation and in dynamical equilibrium \`a la Mitchie-King.  Fragmentation in hydrodynamical calculations often proceeds from filaments and knots  (e.g., \citealt{klessen01,maclow04,maschberger10,bate14}). The clumps that form in a fragmenting 
 Hubble  flow are also surrounded by filaments and other structures which perturb them.

\subsection{Clump mass function} 
\label{Sec:ClumpMassFunction}

\begin{figure*}
\begin{center}
\includegraphics[width=\textwidth]{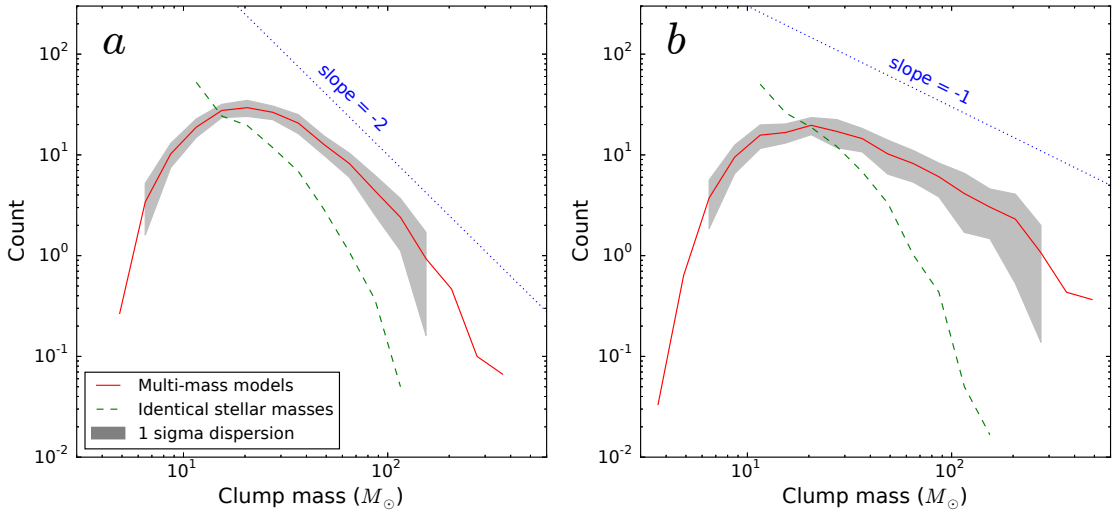}
\caption{Mass function of the clumps identified with the MST algorithm. The calculations all had $N = 15 000$ stars, and we have averaged over 30 realisations for each configuration.  The results for three stellar mass functions  are displayed~: a model with equal-mass stars (green dashed line)~; a Salpeter distribution function truncated at $20 M_\odot$ (solid red line, left)~; a Salpeter distribution function truncated at $100 M_\odot$(solid red line, right). a) The clumps  mass function for equal-mass models  shows a trend with mass roughly in agreement with an $M^{-3}$ power-law. By comparison,  the results for an  Salpeter stellar distribution function  truncated at $20 M_\odot$ has a bell-shaped profile, with a peak around $M = 20.5 M_{\odot}$~; only the tail-end shows marginal agreement with an $\propto M^{-2}$ power-law (dotted line on the figure)~; b) another Salpeter distribution function but with  the upper-mass truncation  set at $100 M_\odot$. The tail at large clump mass is now much flatter, with a slope $\approx M^{-1}$, (dotted line on the figure as well). The bins used  had constant logarithmic mass intervals. } 
\label{Fig:clumpMF}
\end{center}
\end{figure*}

We wish to quantify the relation of the clump- to  the {\it stellar} mass function in the generated initial conditions. To this end, we ran a set of simulations where all the stars have the same mass, and two sets for which a Salpeter mass function with $\alpha = 2.35$  was truncated at different upper- and lower-bounds. A total of 15000 stars in a Hubble configuration were used, all let go  with the same initial expansion rate  $\Hub_o = 1$. For the multi-mass models, the mass range  was chosen as $[0.3, 100]\, M_\odot$ and $[0.35, 20]\, M_\odot$ so that the mean stellar mass $= 1M_\odot$ as for the single-mass models. Thirty different runs were performed in each case and the outcome averaged for better statistics. These are refered to as Rmh1, Rmh100 and Rmh20 in Table \ref{Tab:fragmentationmodels}.

On Fig.~\ref{Fig:clumpMF}, we display the  number of clumps as function of clump mass for the truncated Salpeter  models as a red solid line, while the single stellar mass models are shown in green dash. A grey shade indicates one standard deviation where statistics allow ({\it i.e.}, large numbers), and we have used bins of constant logarithmic mass intervals.  Fig.~\ref{Fig:clumpMF}a shows Rmh20 models, and \ref{Fig:clumpMF}b shows Rmh100 models. 
Note that the number counts gets down to fractionnal values due to the averaging procedure. With clump membership restricted to $N \geq 12$, the identical-mass model stays close to a power law (straight dotted line on the figure) of index $\approx -3$ down to the lowest clump mass of 12$M_\odot$. A spread in stellar masses leads to fewer low-mass clumps forming, which profits the  more massive ones (we counted $\simeq100$ clumps of $12 M_\odot$ for the equal-mass case~; and $\approx 32$ with a mass $ \le 12 M_\odot$ for the other ones) . This transforms the clump mass function, from a near-power-law, to a bell-shaped distribution.  

When very massive stars are included in the calculations, yet more massive clumps are formed (Fig.~\ref{Fig:clumpMF}b). The formation of large sub-structures depletes the number of clumps around the peak value, and so the distribution becomes broader and shallower. The mean clump mass for the different cases read $18 M_\odot$ (equal-mass), $30 M_\odot$ (Salpeter $m_{max} = 20 M_\odot$) and $42 M_\odot$ (Salpeter $m_{max} = 100 M_\odot$), a steady increase with the width of the stellar mass spectrum. On the other hand, the position of the peak of the distribution remains unchanged at (roughly) $20$ to $21M_\odot$. The trend in total number of clumps detected is a slight {\it de}crease with the broadening of the  stellar mass spectrum, from 166, down to 143 respectively for the $m_{max}=20$ and $100 M_\odot$ Salpeter models.  
  We observe that the overall fraction of  stars found in clumps (some $\approx 6500$ out of 15000, or 43\%) stays unchanged.
  
We argue that the shape of the clump mass spectrum provides indirect evidence for the role of massive stars predominantly as seeds for growth in our simulation, as opposed to the hierarchical build-up of  clumps from very tiny sub-structures. There are two tell-tale signs to support this view~: a) if high-mass clumps formed through the repeated and stochastic merger of smaller clumps, then the clump mass function should tend to a log-normal distribution, which is  symmetric (in logarithmic scales) with respect to the peak value, whereas the distributions shown here lack this basic property~; and b) the ratio of maximum clump mass to mean mass may exceed 15 when the  stellar truncation mass is set to $20\, M_\odot$, and reaches only $\sim 4$ in the case when the upper mass is set to $100\, M_\odot$. If small-ish clumps were merging at the same rate in both models, then this ratio should be comparable. Instead, very large clumps take too long to assemble and the merger rate drops with clump mass. Recall that all fragmentation calculations ran for the same total time. 

To check whether massive stars act as seeds in the simulation, we borrow from black hole dynamics in galactic nuclei the notion of a {\it radius of influence}, which is the radius  enclosing as much mass in the stars as the central black hole mass (see e.g. \citealt{merritt13}).  Here, the stars inside the influence radius are bound to the massive star at the centre. Thus if a massive star is a seed for a clump, and only the stars inside the influence radius remain bound to it, we should count as many clumps in the mass range $2m_\star, 2 m_\star + 2{\rm d}m_\star$, as there are stars in the range $m_\star, m_\star + {\rm d}m_\star$.  Because the maximum clump mass exceeds twice that of the most massive stars $m_{max}$, some degree of merging must take place. If we count all clumps starting from the truncation value $m_{max}$ of the stellar mass function,  
then we should find as many clumps in the mass range above $m_{max}$, as there are stars in the interval $[m_{max}/2, m_{max}]$.  We find for runs with $m_{max} = 20 M_\odot$ some $93$ clumps more massive than that, when there are $\simeq 100$ stars in the range $[10, 20] M_\odot$, essentially identical~; and some $11$ clumps of $100 M_\odot$ or more, when there are (on average) 9 stars in the mass range $[50, 100] M_\odot$. This calculation suggests that  most 
massive stars act as seeds for the formation of large clumps in the generated initial conditions.

\subsection{The stellar mass function in clumps} 

\begin{figure}
\includegraphics[width=0.5\textwidth]{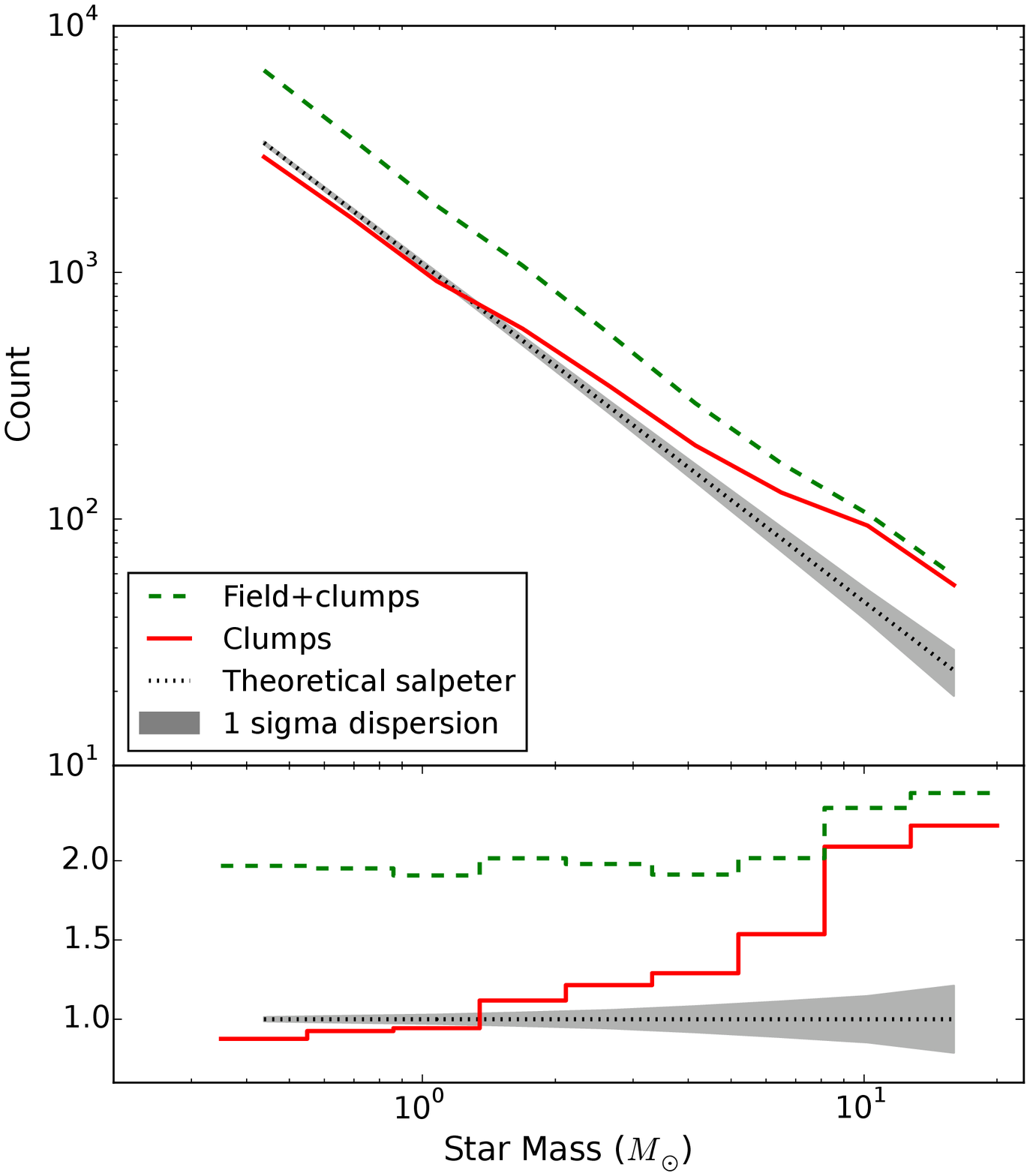}
\caption{Top panel~: Mass function of all stars belonging to a detected clump (solid red). The expectation drawn from a Salpeter distribution function for the same total number of stars in dotted black~; the grey shade are $1\sigma$ uncertainties. The green dashed line is the  distribution for the full cluster. Bottom panel~:  same data normalised to the Salpeter expectation.}
\label{Fig:MFmembers20}
\end{figure}

To complete the analysis, we show on Fig.~\ref{Fig:MFmembers20} the mass function of stars both in clumps and in the whole cluster. For brievety, we only show a model with a mass function truncated at $20 M_\odot$, however our conclusions are not sensitive to the truncation value. The mass function of $\approx 6400$ stars that were found in clumps (some 43\%) is displayed as the red solid curve. The theoretical Salpeter distribution function for the same number of stars is shown in black dots, with grey shades giving the  $1 \sigma$ dispersion from multiple samplings. 
Finally, the green dashed curve  shows the mass distribution of all 15 000 stars in the model. The lower panel is the same data normalised to the Salpeter data. The uptake  in massive stars for the whole population (green dashed line) of both clumps members and field stars is a statistical artifact and lie within the standard deviation of a Salpeter distribution with comparable sampling number. The clump member population clearly deviates from a Salpeter distribution in two ways~: first we note a deficit of low mass stars with respect to the field population (shown as the green dash)~; secondly, although a Salpeter mass function is more or less consistent with the population up to $M\approx 2M_\odot$(black dotted line) the distribution shows a clear excess of massive stars. We find that practically all the stars more massive than 10$M_\odot$ ended up in a clump (this is the point where the solid red curve joins the dash green one). A linear regression fit of the clump members mass function gives a power-law index of $-2.15 \pm 0.02$, shallower than the Salpeter index of -2.35. Applying the same analysis to field stars,  we find a steeper mass function of index $-2.46 \pm 0.02$. 
The difference of $\approx 0.3 $ between the two populations is very similar to what is found in the Milky Way disc (see e.g. 
\citealt{czekaj14,rybizki15,bastian10} )

\begin{figure}
\includegraphics[width=0.5\textwidth]{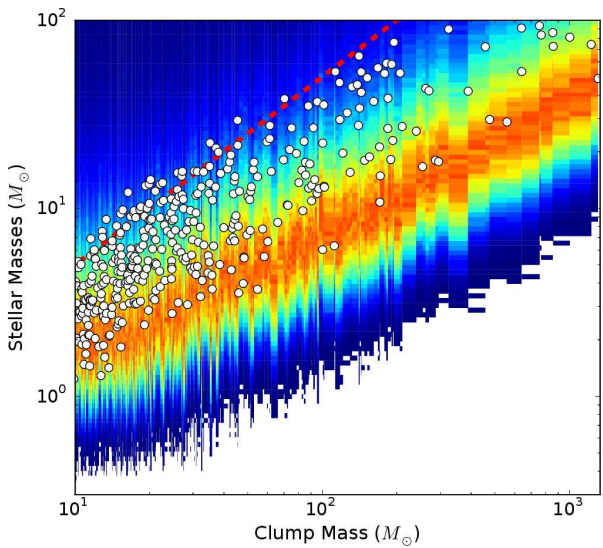}
\caption{Mass of the heavier star in each clump, shown as white dots, as a function of clump mass. The color map shows the likelihood for the maximum mass if all clump members were  sampled from a Salpeter IMF~; the orange crest gives the maximum likelihood. The red dashed line shows the  relation $m_{clump} = 2 m_{max}$ (see. \S\ref{Sec:ClumpMassFunction}). The data was taken from the R40h100 run.}
\label{Fig:MaxMass}
\end{figure}

\cite{bonnell04} and \cite{maschberger10} showed from inspection of  hydrodynamical simulations that massive stars play a key role in 
the assembling process of clumps, attracting already formed protostars to them. We find  a similar general trend in Hubble-fragmented gas-free simulations: clumps develop around massive stars so that their stellar mass function is top-heavy. This excess can also be seen in the top panel of Fig.~\ref{Fig:MaxMass} in which for each of 440 clumps, we show as white dots the mass of their heaviest star as a function of the host clump's mass. These data were obtained from the R40h100 run (cf. Table 2). For comparison, we sampled a Salpeter mass function, drawing the same number of stars as found in each clump. We then identify the most massive star in the Salpeter sample~; the procedure was repeated 15000 times {\it for each clump} to obtain suitable statistics. The grey shades (color levels in the electronic version) 
 shows the resulting distribution. In a nutshell, Fig.~\ref{Fig:MaxMass} shows for each clump the likelihood that their most massive stars may be drawn from a Salpeter function. Only clumps with a mass $> 10 M_\odot$ are included to account for a possible bias when clump membership reaches below $N_d =12 $ stars. It can be seen on the figure that the scatter of white dots tends to lie systematically above the crest of maximum likelihood (light shade on the figure). If we add the  relation $m_{clump} = 2\max\{m_\star\}$ (cf. \S5.3), we find some overlap with the data (see the red dashed line on Fig.~\ref{Fig:MaxMass}). This clearly illustrates the tendency for massive stars to act as seeds when the clump form, while the scatter is driven by the merger and accretion history of individual clumps. 
The correlation displayed on Fig.~8 is in good agreement with observational data for young embedded clusters of the same mass range  (see \citealt{weidner13}). Note how the {\it scatter} in the correlation brought on by the dynamical processes at play during the adiabatic fragmentation phase also compares well with the data. Thus the stellar clumps modelled here recover an important characteristic of observed embedded young clusters.


\subsection{Clump mass segregation}
\label{Subsec:ClumpSegregation}
In this section, we ask whether the clump assembling process at play in our simulations accounts for the mass segregation measured  in  star-forming cores in hydrodynamical simulations. The measure of mass segregation of \cite{olczak11} based on the MST, while efficient, will give noisy results for very small-N clumps. Instead, we follow \cite{maschberger10} and rank clump members according to their distance to the geometric centre of a clump, which is calculated by number-averaging (so this centre is not the clump barycentre). We then sort the bodies by mass and tabulate the radial rank of the three most massive ones. The great advantage of this approach is that it is independent of geometry and absolute size once the ranking is normalised to the clump membership $N_c$.  One issue arises with the binning of the rank, since small values of $N_c$ give large intervals by construction, and conversely for populous clumps~: we found a good compromise by setting the width of each bin to 1/20 since the mean clump mass $\sim 20 M_\odot$ implies $N_c \sim 20$ on average. The procedure is repeated over all clumps identified in the run (typically on the order of $\sim 200$). The diagnostic for an un-biased sampling is a profile with radius that remains the same regardless of the mass selected~; if, furthermore, the stars are (on the mean) un-segregated in radius, then the profiles will be flat. 

\begin{figure}
\includegraphics[width=0.5\textwidth]{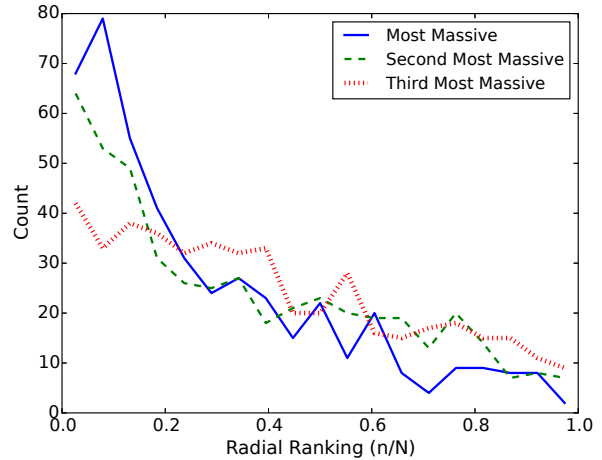}
\caption{Histograms of radial ranking of first, second and third most massive star in each clump for a model with N = 40 000 stars (R40h100).}
\label{Fig:clumpMseg}
\end{figure}

Fig.~\ref{Fig:clumpMseg} graphs the  distribution of rank of the three most massive stars in all the clumps from R40h100 fragmented state. The salient features are that 1) none of the distributions is flat, all three peaking significantly  at small ranks~; and 2) there is a clear trend for the most massive star also to be the most  segregated. Precisely this result had to be expected from the internal dynamics of small clumps (cf. \S3.2).  Our Fig.~\ref{Fig:clumpMseg} should be compared with Fig.~13 of \cite{maschberger10}: it is striking that the measure of mass segregation attained here for a gas-free configuration is a good match to a full hydrodynamical setup. By implication the segregation proceeds more vigourously once the proto-stellar cores have condensed and behave essentially like point sources. The initial configuration that we have adopted relies only on Poisson noise to seed clumps, however once again we find evidence that massive stars begin and remain the centre of gravitational focus for clump formation. That is not so when clumps are setup using a fractal approach \citep{goodwin04,allison09}. There is then no segregation initially, and it all develops at or shortly prior to the global system evolution towards equilibrium (the violent relaxation phase). 

In the next section we follow through with the final stage of evolution towards equilibrium and compare the final configuration with those of \cite{allison09} and the recent study by \cite{caputo14}. 

\section{Evolution towards equilibrium}
\label{Sec:collapse}

\begin{figure}
\begin{center}
\includegraphics[width=\columnwidth]{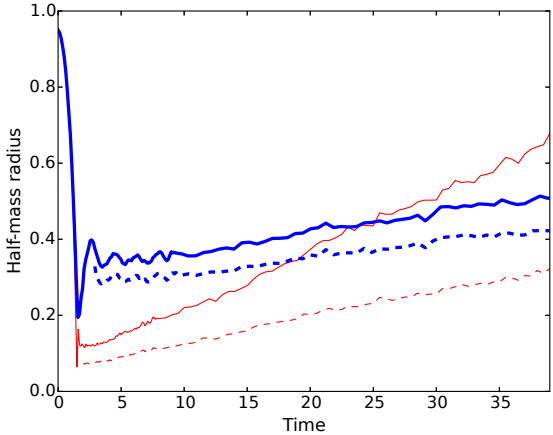}
\caption{Half-mass radius as function of time for two systems undergoing collapse~: a uniform-density sphere (thin red solid curve) and a clumpy Hubble model (thick blue solid curve). Half-mass radii are in H.u, as well as the time axis, where $t_{Henon} = 1 {\rm unit} =  0.270 \times 10^5 $ years. Dashed lines are the half-mass radii of the same systems for the same systems but including only the bound stars.}
\label{Fig:Rhm_global}
\end{center}
\end{figure}

\begin{table}
\begin{center}
\caption{Summary of collapse simulations and their characteristics. These simulations started from a subvirial state: cold uniform sphere or fully fragmented Hubble model; each were evolved up to t = 40 H.u}
\label{Tab:evolution_models}
\begin{tabularx}{\columnwidth}{XXlXX}
\hline
Name & N & Mass range & Model \\
\hline
Rh100 & 15000 & [0.3 - 100] & Hubble  \\
Rh20 & 15000 & [0.35- 20 ] & Hubble  \\
Ru100 & 15000 & [0.3 - 100] & Uniform  \\
Ru20 & 15000 & [0.35- 20 ] & Uniform  \\
\hline
\end{tabularx}
\end{center}
\end{table}

The Hubble expansion comes to a halt at a time $\tau$ when $\theta(\tau) = \pi/2$  in Eq. (\ref{Eqn:Expansion}); (see Appendix \ref{App:Hubble} for details). The system as a whole is then in a sub-virial state. We wish to explore briefly the  violent relaxation that follows and the equilibrium that ensues. In the present section, simulations will use the fully fragmented state of Hubble models as initial conditions for the subsequent dynamical evolution. Observationnal clues point to  collapsing and violently relaxing clusters. For example, \cite{cottaar15} find IC348, a young (2-6 Myr) cluster, to be both super-virial and with a convergent velocity field, consistent with infalling motion. Dry collapse with no gas is an idealized situation: clearly if there was residual gas between the clumps and it was evacuated through stellar feedback, both the clump merger rate and the depth of the potential achieved during relaxation would be affected. As a limiting case, rapid gas removal may lead to total dissolution (see for instance \citealt{moeckel12} and \citealt{fujii16}). In the current situation, all clumps will merge.

The numerical integration were done once more with the Nbody6 integrator with the same computational units. For comparison purposes, we also performed simulations of cold uniform sphere, a configuration which has been extensively used  in the past (e.g.,\citealt{theis99,boily02,barnes09,caputo14,benhaiem15}) and one that minimises the level of fragmentation and mass segregation in the on-set of collapse. Table~\ref{Tab:evolution_models} lists the simulations performed. We focus here on models with a mass function from 0.35$M_\odot$ to 20$M_\odot$ and 15000 stars, a compromise value for rich open clusters that should allow us to identify clearly collisional effects and trends with time, and ease comparison with the recent study by Caputo et al. (2014) where most calculations are performed with that sampling. We let both Hubble-fragmented- and uniform sphere evolve up to 40 H.u. 

\subsection{Scaling to physical units}
Before discussing the results, it is useful to translate the units of computation to physical scales. This is important if we want to discuss the state of the systems using one and the same physical time, such that the hypothesis of no stellar evolution holds good.
This complicates the analysis, because all the time-scales defined in \S 3.1 are based on the hypothesis of equilibrium, and we start-off out of equilibrium. To make things clearer, let us resize the configurations so that the half-mass radius $r_h = 1$ pc initially, with a total system mass of $M = 15\times 10^3\, M_\odot$ for a volume density $\rho = M / (8\pi/3 r_h^3) \simeq 10^3 M_\odot / pc^3$, well within values typically inferred from observations. 
We then compute a free-fall time for the uniform-density model of $t_{ff} = \pi /2 \left( r_h^3 / GM \right)^{1/2} \approx (\pi/24)^{1/2} \times 10^5 $ years~; the {\it initial} crossing time would then be $t_{cr} = 2 r_h / \sigma_{1d} = 4\sqrt{6}/\pi \,t_{ff} $ where we invoked the virial theorem to compute $\sigma^2_{1d} = 1/3 \times GM/2 / r_h $. (The crossing time is larger than $t_{ff}$ because all the mass goes into the origin during free-fall.)  
In practice the more useful crossing time has to be computed from the equilibrium state achieved. If we once more invoke the virial theorem and note that at constant mass and energy the equilibrium state would reach a size $r_h^{eq} \approx r_h^{(0)}/2$ or half the initial radius, then one computes $t_{cr}^{eq} \approx t_{cr}^{(0)} / 2\sqrt{2} =  2\sqrt{3}/\pi t_{ff} \approx 1.10 \, t_{ff}$. 
Direct computation of the problem of a collapsing sphere give $t_{ff} \approx 1.36$ H.u. We therefore use the conversion factor \[ \frac{t}{{10^5\,{\rm years}}} = (4/3) / (\pi/24)^{1/2} \simeq  3.7\, t_{Henon} \,. \]  The equilibrium crossing time is then $t_{cr}^{eq} = 1.1\, t_{ff} = 3/2 \, t_{Henon} \approx 0.4 \times 10^5 $ years. The time-conversion factor adopted is conservative and does not factor in the stars that may escape during virialisation.
Thus by running up to 40 H.u we ensure that the systems evolve for at least 25 crossing times and $ \sim 10^6 $ years, short before the lifetime of massive stars.  With $N = 15 000$ and a mass range of $m_{max}/\langle m\rangle = 20$ we find from (\ref{Eqn:Trel}) and (\ref{Eqn:Tms}) a two-body relaxation time of $t_{rel} \approx 80\,t_{cr} $ (120 $t_{Henon}$, or 3 Myrs) and mass-segregation timescale of $t_{ms} \approx 4\, t_{cr} $ ($6\, t_{Henon}$, or $1.6\times 10^5$ years).  
  
\subsection{Collapse and virialisation}
\label{Subsec:Collapse}
The constant diffusion of kinetic energy by two-body interaction means that no stellar system ever reaches a steady equilibrium. However we can contrast the time-evolution of two configurations and draw conclusions about their observable properties. 

\begin{table}
\caption{Number of initially ejected stars in two collapse calculations} \label{tab:Ejectedstars}
\begin{tabular}{lrr}
Run & Ejected stars & Ejected mass  \\
\hline
Ru20  &  4227 & 27\% \\
Rh20  &  1932 & 12\% \\
\hline
\end{tabular}
\end{table}

With this in mind  we turn to Fig.~\ref{Fig:Rhm_global} in which we show the evolution of the half-mass radius for the cold uniform model (labeled Ru20~; thick red curve), and the Hubble model (labeled Rh20~; thin blue curve). Both systems have the same bounding radius initially, contract to a small radius when $t \simeq 1.4 $ units and then rebound at time $t \simeq 2 $ units. When all the stars are included in the calculation for $r_h$, we find that the radius increases at near-constant speed after the collapse. That trend does not appear to be slowing down which indicates that a fraction of the stars are escaping. The first batch of escapers is driven by the violent relaxation, however the trend continues beyond $ t = 10$ units, corresponding to $t > t_{ms}$ which implies two-body scattering and effective energy exchange between the stars. Note how the uniform model has a much deeper collapse and rebounds much more violently, shedding a fraction twice as large of its stars (Table~\ref{tab:Ejectedstars}). The half-mass radius $r_h$ increases steadily in both models, from the bounce at $t \approx 2$, until the end of the simulation (values in H.u):  

\begin{tabular}{lllrr}
$r_h$ Uniform & 0.11 &  $\rightarrow$ & 0.63 & ($\times 5$); \\
$r_h$ Hubble & 0.34  &  $\rightarrow$ & 0.49 & ($\times 1.4$). \\
\end{tabular}\\

%
Clearly the gentler collapse of the fragmented model has led to a more extended post-collapse configuration and reduced two-body evolution. Observe how the uniform model Ru20 is ejecting more stars than the Hubble model~: if we repeat the calculation for the Hubble run Rh20 but now include only the bound stars\footnote{See Appendix A for details of the selection procedure.}, the curve of $r_h$ obtained and shown as dash is shifted down but keeps essentially the same slope $\approx  0.004$. By contrast, the calculation for the bound stars of run Ru20 yields a much shallower slope than for the whole system: the slope drops from 0.015 to about 0.007. Irrespective of how the half-mass radius is calculated, the conclusion remains the same and agrees overall with the remark by \cite{caputo14} that boosting the kinetic energy of the collapsing initial configuration softens the collapse~; this was shown in a different context by \cite{theis99} and confirms these older findings.  Here, the fragmented model has finite kinetic energy due to the clumps' internal motion. The important new feature brought by the fragmented initial conditions is that the {\it mass profile} of the virialised configuration evolves much less over time in comparison. 


\begin{figure*}
\begin{center}
\includegraphics[width=0.75\textwidth,clip=true]{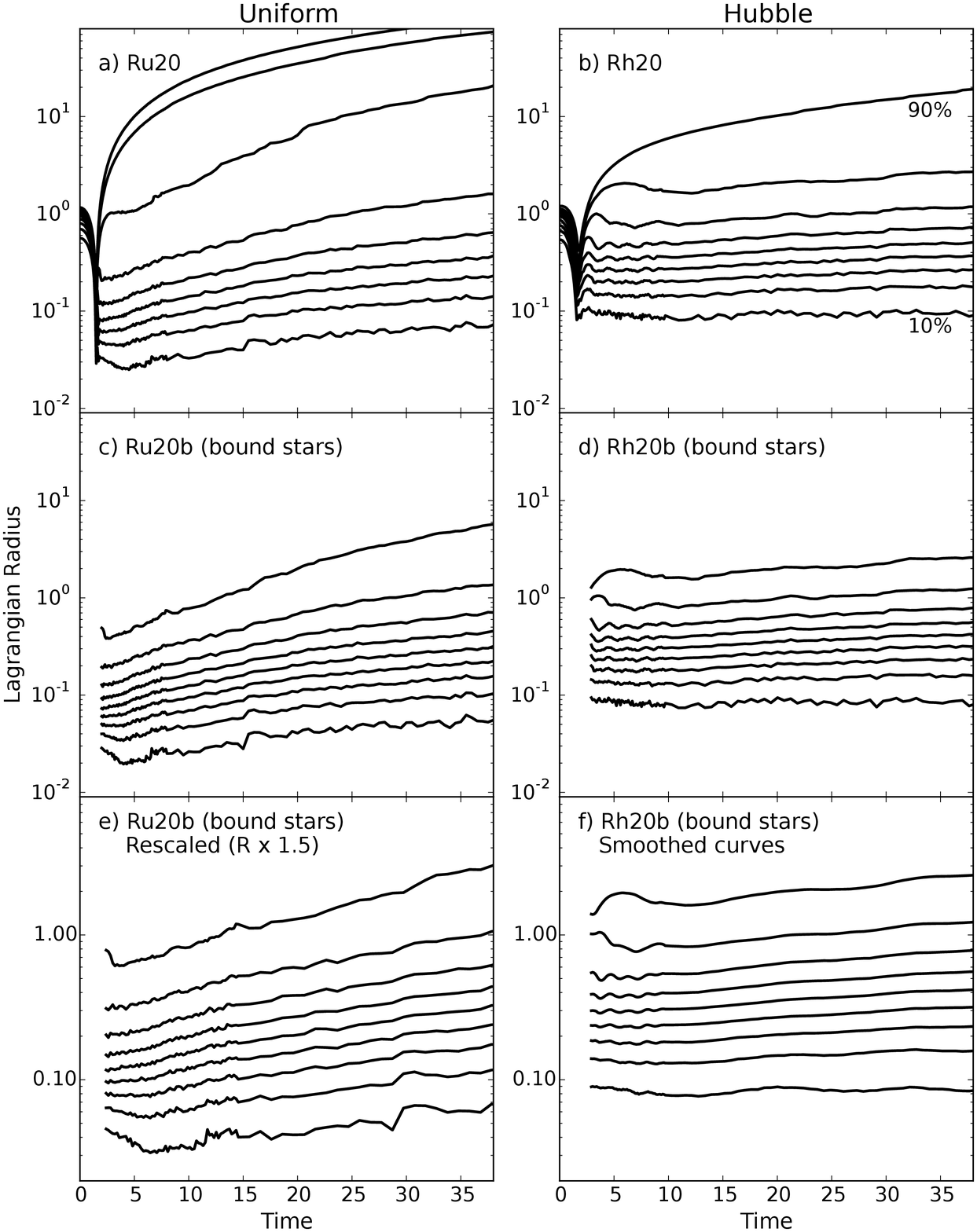}
\caption{The ten-percentile mass radii (10\% to 90\%) as function of time. Radii and time axis are in H.u, with $t_{Henon} = 1 {\rm unit} =  0.270 \times 10^5 $ years. Left panels show the Uniform model and right panels show the Hubble fragmented models. Panels a and b show the evolution of the whole systems, while panels c and d shows the same radii computed for the bound stars only. Panel e shows the Uniform bound model (Ru20b) for which radius and time were rescaled to compensate the difference of initial kinetic energy (see text for details). Panel f shows the same information as panel d with smoothed data. 10\% and 90\% radii are labelled in the top right panel.
}
\label{Fig:Lagrange}
\end{center}
\end{figure*}

At the bounce, the half-mass radius of the Hubble model is $\approx 4$ times larger than that of the of the initially uniform sphere at rest (Fig. 10). The half-mass radii overlap at time $t \approx 15 \,H.u.$ (solid curves) or $t \approx 50 \,H.u.$ (dashed curves). Is the same trend applicable to all Lagrangian radii? To answer this question we plot on Fig. 11 the ten-percentile mass radii for the two models. The results are displayed for the two situations including all the stars (top row) or bound stars only (middle row). It is striking that the curves show very little evolution at all mass fractions for the case of the Hubble model (see right-hand panels on the figure), whereas all mass shells either contract or expand in time for the uniform one. We have noted how this model undergoes two-body relaxation on a timescale of $t \approx 10 \,H.u$: the innermost 10\% mass shell shows an  indication of \textit{core-collapse} at $t \simeq 5 \,H.u.$. We note here that the two sets of curves reach very similar values at the end of the calculations ($t = 40\, H.u$). A key difference between the two models, therefore, is that the final configuration of the Hubble model is almost identical to what it was at the bounce ; the same simply does not hold in the case of a uniform-density collapse. Furthermore, the Hubble calculation shows no hint of two-body relaxation or core-collapse. This raises the possibility that the system properties in the final configuration remain better correlated with those at the on-set of (global) collapse (we return to this point in \S7).

\cite{caputo14} and \cite{theis99} noted how a finite amount of kinetic energy in the {\it initial} configuration alters the  depth of the bounce during collapse. The ratio of half-mass radius at the bounce, to its initial value, is then $ r_h/r_h(0) \simeq Q_o + N^{-1/3}$, where $Q_o$ is the virial ratio of the initial configuration (see Fig. 5 of Caputo et al. 2014). We computed the kinetic energy of the  Hubble configuraiton and found that the internal motion of the clumps means that $Q_o (Hubble) \simeq 0.02$ for a Salpeter mass function with upper truncation value of $20 M_\odot$. With $N = 15k$ stars, the ratio $r_h/r_h(0) \simeq 0.041$ when $Q_o = 0$ shifts to $r_h/r_h(0) \simeq 0.061$ when $Q_o = 0.02$, or a factor close to 3/2. To account for the difference in kinetic energy of the initial configurations, we may therefore rescale the uniform model such that positions are  $ \times 3/2$ and the time unit is $\times (3/2)^{3/2} \simeq 1.84$. The new configuration would evolve in time in exactly the same way after mapping positions and time to their rescaled values. The result is shown as the bottom row on Fig. 11.  Note that we  have blown up the vertical axis to ease comparison between uniform- and Hubble models with bound stars only included. The rescaled uniform model is now slighlty more extended than before, but overal the final two configurations (at $t = 40 \,H.u.$) are as close as before rescaling. This demonstrates that  the outcome of the uniform collapse and its comparison with the Hubble model is not sensitive to a small amount of initial kinetic energy. We note that while the ratio $Q_o$ is a free parameter in many setups for collapse calculations, that parameter is fixed internally in the Hubble approach.

\section{Global mass segregation}
\label{Sec:Segregation}

To investigate the state of mass segregation in our models, we follow the analysis of Caputo, de Vries \& Portegies Zwart (2014). The masses are sorted by decreasing values, then subdivided into ten equal-mass bins. This means that the first bin contains the most massive stars. The number of stars in each bin increases as we shift to the following bins, since their mean mass {\it de}creases, and so on until we have binned all the stars. The half-mass radius $r_h$ computed for each bin is then plotted as function of time. In this way the mass segregation unfolds over time: if the stars were not segregated by mass, all radii $r_h$ would overlap. If two sub-populations share the same spatial distribution, their respective $r_h$ will overlap.

\begin{figure*}
\begin{center}
\includegraphics[width=0.75\textwidth,clip=true]{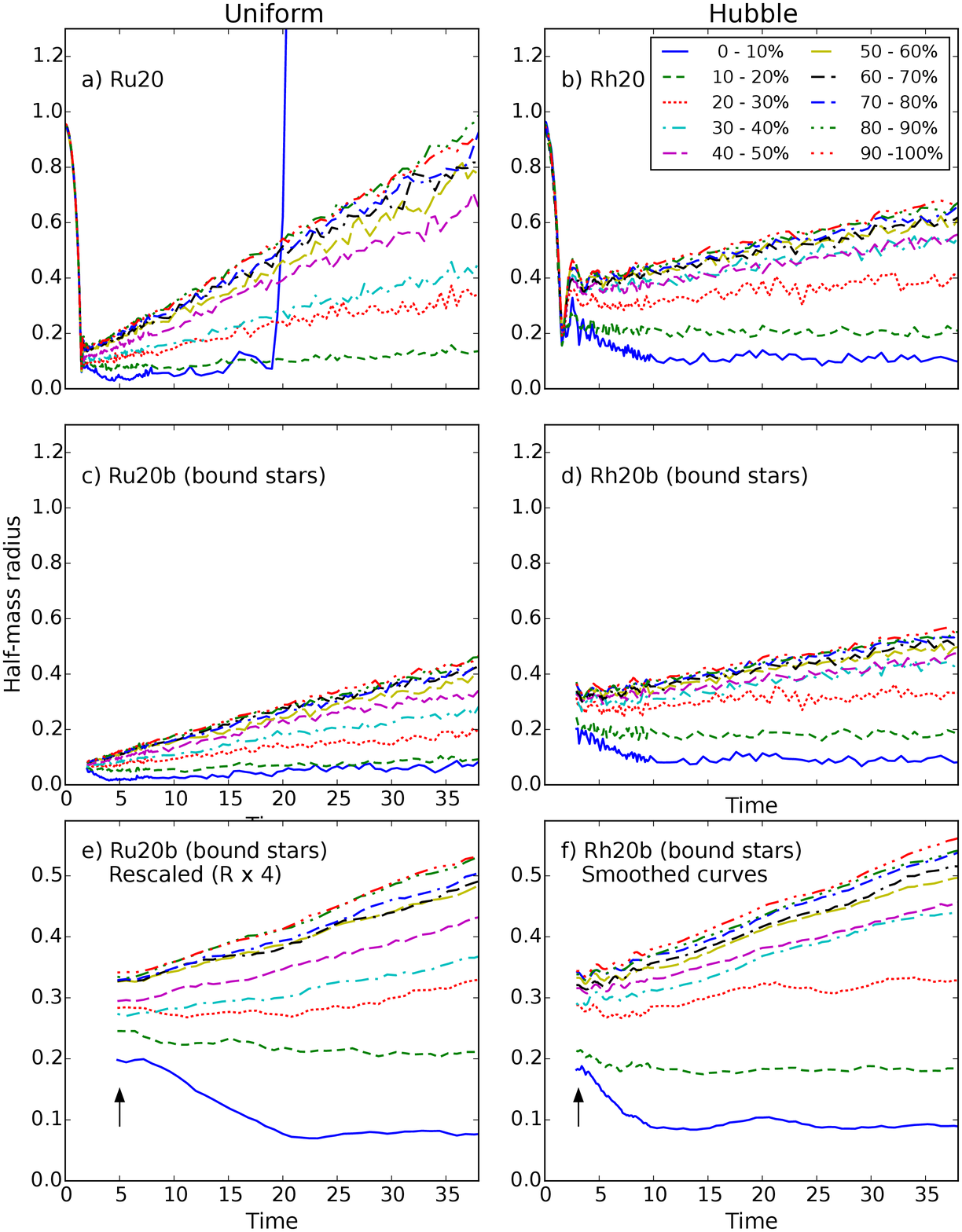}
\caption{Half-mass radii of stars selected by mass as function of time. Each bin identified with 0-10\%, 10-20\% .. 90-100\%, contains ten percent of the total system mass. The stars where sorted by mass in decreasing order, and used to fill each ten-percent mass bin in order. Hence the first ten-percentile contains the most massive stars, the next ten-percentile the second group of massive stars, and so on until the 90-percent bin which contains the least massive stars in the model and is the most populated. Half-mass radius and time are in H.u, with $t_{Henon} = 1\,unit = 0.270 \times 10^5$ years. Left panels show the evolution of the Uniform model (Ru20, Ru20b) and right panels do the same for the Hubble model (Rh20, Rh20b). The organization of panels follows the same layout than figure~\ref{Fig:Lagrange} with a different factor for the rescaling of the uniform system. }
\label{Fig:Rhm_segr}
\end{center}
\end{figure*}

Figure \ref{Fig:Rhm_segr} graphs the results for initially uniform-density- and fragmented Hubble models. The layout of the figure is the same as for Fig. 11. The violent relaxation phase leads to mass loss for both models and the much more rapid expansion of the half-mass radii of low-mass stars is an indication that most escapers have a lower value of mass. Fig.~\ref{Fig:Rhm_segr}(c) and (d) graphs $r_h$ for the bound stars of each sub-population. Clearly the initially uniform-density model is more compact early on, but note how the heavy stars sink rapidly to the centre, more so than for the case of the Hubble model. The spread of half-mass radii increases with time for both models, however two-body relaxation in the uniform-collapse calculation is much stronger, so that by the end of the simulations the half-mass radii of the low-mass stars of the respective models are essentially identical. Since the low-mass stars carry the bulk of the mass, that means that the two models achieve the same or similar mean surface density by the end of the run. At that time, the heavy stars in the uniform-collapse calculation are clearly more concentrated than in the Hubble run (compare the radii out to $\sim 40\%$ most massive stars). A direct consequence of this is that the {\it color} gradients of the core region of a cluster are much reduced when the assembly history proceeds hierarchically, in comparison with the monolithic collapse. It will be interesting and possibly important in future to compare such models with actual data for young clusters.

Another interesting remark is that the kinematics of the stars within the {\it system} half-mass radius is much different between the two models. For the Hubble calculation, the system half-mass radius, $ \approx 0.43 $ H.u, at $t = 40$ (cf. Fig. 11[d]) coincides with the half-mass radius of the $30-40\%$ bin stellar sub-population. All bins up to that range show little or no time-evolution, around the end of the run, which we interpret as efficient retention of these stars by the relaxed cluster. In the case of the uniform-collapse run, the system half-mass radius reaches $\approx 0.33$ H.u., which is significantly larger than the radius for the $30-40\%$ stellar sub-population. For that model, only the bins $0-10\%$ and $10-20\%$ are flat, and all the others increase almost linearly with time. Thus a fair fraction of bright stars deep in the cluster show systematic {\it outward streaming} motion, along with low-mass ones. This brings up the possibility to measure this signature motion through relatively bright stars, originating well inside the cluster half-mass radius. Recall that only post-bounce bound stars where selected to compute $r_h$ on Fig.~\ref{Fig:Rhm_segr}(c) and (d) ; the expansion is therefore not driven by chance events (e.g., Fig.~\ref{Fig:Rhm_segr}[a]), but rather through two-body relaxation. On the down side the bright tracers would be short-lived, and this may prove a strong constraint for observational detection.

Given the early dynamical evolution associated with substructured stellar clusters, some observed dense objects may yet be out of equilibrium (see \S\ref{Sec:discussion}). We wish to investigate the out-of-equilibrium state of our models just after the collapse. To ease the comparison between the two systems, the same rescaling procedure as for Fig~\ref{Fig:Lagrange} was applied to the uniform model, only this time the scaling was chosen so that the two clusters have comparable densities after the bounce. Lengths were multiplied by $4$; the time-axis is then scaled up by a factor $(4)^{3/2} = 8$. The result can be seen in panel (e); panel (f) shows a smoothed and zoomed in Hubble model for comparison.

\begin{table*}
\caption{Values of half-mass radii and their ratio to that of the most massive stars. The results are for the rescaled bound uniform model (rescaled Ru20b) and the bound Hubble model (Rh20b), after the collapse, and before dynamical mass segregation sets in.} \label{Tab:RhmVal}
\begin{tabular}{l|llllllllll}
Uniform (rescaled) & 0-10\% & 10-20\% & 20-30\% & 30-40\% & 40-50\% & 50-60\% & 60-70\% & 70-80\% & 80-90\% & 90-100\% \\
\hline
Radius   & 0.20 & 0.245 & 0.282 & 0.273 & 0.294 & 0.325 & 0.326 &  0.328 & 0.335 & 0.340 \\
Ratio    & 1.0 & 1.23 & 1.41 & 1.37 & 1.47  & 1.63 & 1.63 &  1.64 & 1.68 & 1.70 \\

\hline
Hubble  & 0-10\% & 10-20\% & 20-30\% & 30-40\% & 40-50\% & 50-60\% & 60-70\% & 70-80\% & 80-90\% & 90-100\% \\
\hline
Radius  &  0.18 & 0.21 & 0.286 & 0.293 & 0.316 & 0.321 & 0.333 & 0.338 & 0.342 & 0.344 \\
 Ratio       & 1.00 & 1.16 & 1.58 & 1.63 & 1.76  & 1.78  & 1.85 &  1.88 & 1.90 &  1.91 \\
\end{tabular}
\end{table*}

We compare the values of the different half-mass radii of the various population before the dynamical mass segregation sets in. This process is clearly visible as the drop of the half-mass radius of the most massive stars during the evolution. We are interested in the segregation which originates from the collapse and is present before this dynamical evolution. Table~\ref{Tab:RhmVal} sums up the values of the half-mass radii taken at $t\sim5$ for both models, both corresponding to the same unevolved post-collapse state (see arrows on panels [e] and [f] on Fig~\ref{Fig:Rhm_segr}). With on the order of $\sim 100$ stars per bin or more, one estimates roughly a ten-percent standard deviation from random sampling. To measure the \textit{relative} segregation between populations, the table also lists the ratios of each half-mass radius to the one for the most massive stars. Both models appear mass segregated (since these ratios are significantly greater than unity). The Hubble model is more segregated, on the whole, albeit in a different way compared to the uniform model. The segregation in that one is more regular and spreads over more mass bins. In the Hubble model, the segregation is much enhanced for the first two mass bins. Such differences in the degree and nature of segregation can be explained by the clumps structure before the collapse. We showed in \S\ref{Subsec:ClumpSegregation} the clumps were mass segregated with the most massive members being preferentially located at their center. The low membership and mass of most clumps implies that segregation mostly affects the very top of the stellar mass function. This segregation, predominant among massive stars, is then found in the resulting centrally concentrated system, after the collapse, and visible on Fig.~\ref{Fig:Rhm_segr}. The inheritance of mass segregation was studied by \cite{mcmillan07} for the case of merging Plummer spheres. \cite{allison10} furthermore showed that mass segregation in the system as a whole is enhanced for more filamentary  fractal initial condition (lower dimension, $D$ ; see their Fig. 5). Here our results confirm this observation. Mass segregation is a sensitive function of the initial clumpiness of the system and has immediate bearing on the dynamics of the virialised configuration, since all massive stars are more concentrated in the core.

\section{Summary \& Discussion}
\label{Sec:discussion}

We have developed a new approach based on adiabatic fragmentation to set up self-consistent configurations for stellar dynamics that link up the velocity field of stars to their irregular space configuration of arbitrary geometry, such as knots or filaments. The method offers great advantages: it is easy to implement; it can treat an arbitrary number of stars without any resolution issue; furthermore, the level of fragmentation can be tuned through the Hubble parameter. The light computational load allows for statistical ensemble averaging over large samples (see e.g. \S \ref{Sec:ClumpMassFunction}). For instance, the computation time on a single card for 80,000 stars is about 12 hours. The methods has its limitations: the most significant one is the failure to include hydrodynamical effects. Other approaches are based partly on hydrodynamics: such hybrid methods have been successful but remain limited in scope, for instance \cite{moeckel12}, or demanding in computational resources (and so constrain the number of realizations), as in \cite{fujii16}. A future version of the Hubble method should also include hydrodynamical features.

During the fragmentation process, heavy stars act as seeds for the growth of stellar clumps, and so the stellar clumps mass spectrum is shaped in part by the mass function of the {\it stars} that form in the whole star-forming volume. Although the fragmentation through gravitation only does not include the detailed physics of star formation, we noted that hydrodynamical calculations including gas pressure and turbulence suggest that the gravitational potential of massive stars attract more gas and stars and, as such, act act as seeds for the formation of clumps \citep{bonnell04}. We therefore recover a key prediction of hydrodynamical simulations. It is then interesting to ask whether observations show a correlation between the host clump mass and its population of massive stars.

\subsection{Link with observational data}
 Based on analysis of our fragmented Hubble models, we recover on Fig~\ref{Fig:MaxMass} a correlation between the maximum mass of a star found in a clump of a given mass, $M_c$. This $\max\{m_\star\} - M_c $ relation is eerily similar to the compilation for young clusters by \citet[Fig 1, panel C]{weidner13}. Furthermore, we also found that the stellar mass function in clumps has a much flatter (top-heavy) profile than in the field, {\it i.e.} stars that do not belong to any clump: power-law fits for the two stellar populations show that the Salpeter index for clumps stars is lower by about $\approx 0.3$ compared to the same index for field stars. A similar difference is deduced for Milky Way data ( \citealt{czekaj14,rybizki15}; see also Fig 2 from \citealt{bastian10}):  we argue that these characteristics will help tighten our understanding of the long-term evolution of such stellar associations, given that their properties are, on the out-set, close to actual data for young clusters. It should be emphasised that 
 the global index of external galaxies may differ signicantly from the canonical value $\alpha = 2.35$ ({\it e.g.}, the GAMA survey, \citealt{gunawardhana11}; see also \citealt{hoversten08}). We have not explored here to what extent this difference in indices  between field and clump populations will change for other values of the global index $\alpha$.

We have also noted that the clumps are {\it mass segregated} at birth, i.e. at the end of the fragmentation process. When we apply the same ranking statistics as for hydrodynamical calculations of star formation, we obtain the same level of segregation for the three most massive stars in a clump (cf. Fig. \ref{Fig:clumpMseg}). Heavy or light stars caught in dense clumps have high velocities, while  only a small fraction of field stars have such large velocities: we found that the velocity distribution function of field stars is well fitted with a Gaussian, except for the $\sim 1\%$ with the most extreme velocities. We drew a comparison with the SPH calculations by \cite{klessen00}, who attributed the high-velocity tail of gas particles to their in-falling onto stellar clumps. However, we could not identify unequivocally the origin of the large velocities for the field stars (past star-star interactions, attraction by a clump, .. ). That point may well be worth exploring in a future study, as it relates to the likelihood of accretion of field stars by a dense stellar clumps. Recent SPH calculations by M. Bate and collaborators hint at continued exchanges between stellar cores and their environment.

Recent observations of star forming regions are already giving indications how to improve on the adiabatic-cooling approach of this paper. First, \cite{rathborne15} report new ALMA data of the molecular cloud G0.253+0.016, which they show is on the verge of undergoing a burst of star formation. The low sound speed in the $\sim 10K$ gas implies that the proto-stars will condense from the gas and be distributed spatially in a pattern of filaments similar to what is seen in the gas. In the same vein, deep IR observations of $\rho$Ophiucus by \cite{andre07} reveal pre-stellar clumps of cold gas with low inter-clump velocity dispersion, also making a case  for {\it in-situ} star formation. Strong interactions between stars would still impact on the global dynamics but only during the final stages of their formation (binarity, masses of circum-stellar discs, ..).

Finally, the In-Sync survey of \cite{foster15} published APOGEE spectroscopic observations of NGC 1333, a young embedded nearby open star cluster ( $\sim$ 250 pc; total mass of gas + stars $\sim10^3 M_\odot$). The $< 3 $ Myr-old main sequence stars in NGC 1333 have a 1d velocity dispersion $\sim$ 0.8 km/s which matches the expected virial dispersion given the radial mass profile. The stars are surrounded by dense, cool gaseous cores of {\it low} (sub-virial) velocity dispersions. Inspection of the spatial distribution of both the stars and the gaseous clumps show them to be highly substructured (see their Fig. 1). There is an obvious challenge here, discussed at length by \citeauthor{foster15}, to explain why gas-clumps and stars should follow such remarkably different kinematics. To address the transition from embedded clusters to gas-free stellar cores is beyond reach here, and to progress in the right direction will require to include a substantial amount of gas in our scheme before we can explore closer ties with observations of young star forming regions.






\subsection{Clump mass function}

\cite{klessen00,klessen01} fit the gas clump mass function of their simulations with a power-law of index -3/2. On the other hand, the \textit{cluster} mass function in the Milky Way can be described as a power law $\frac{dN}{dM} \propto M^{-\beta}$ where $\beta$ takes value ranging from -2 to -2.4 \citep{haas10}. We have indicated that a power-law relation with a slope $ \beta \simeq -3$ is a rough fit for the case where all the stars are identical (Fig.~\ref{Fig:clumpMF}). This is not so  when a stellar mass spectrum is included~: if a Salpeter distribution function is truncated at $20 M_\odot$ a power-law with slope $ \beta \simeq -2.$ still fits approximately the distribution of clumps of  mass $ > 20\, M_\odot$~; and when the Salpeter distribution function is truncated at $100 M_\odot$, a power-law similarly fits the tail-end of the distribution but now with a slope of $\simeq -1$ (see Fig.~\ref{Fig:clumpMF}b). It is intriguing that the slope of the distribution should fall within the bracket of values for the observational data for clusters ($-2.4^+$) and hydrodynamical fragmentation models (-3/2). If the clumps formed from hydrodynamical fragmentation should become individual star clusters, and recover the $\beta \simeq -2 $ or lower slope of observational 
data, then the distribution function must become steeper and also cover a broader range of masses. The same conclusion applies to the Hubble clump distribution function.

This implies either that clumps will merge so that a few very massive clusters will emerge, or that fewer massive clumps form in the first place. Comparison with existing cluster population needs us to assume these clumps do not fall back and merge through collapse. This is possible with an adequate galactic tidal field ripping apart this fragmented configuration and isolating the clumps before the collapse. Many of the small stellar overdensities detected as clumps would not survive more than a few million years before dispersing through dynamical interaction, however the larger clumps could survive and appear as isolated cluster or part of an association. These massive clumps   are the key to comparison to the galactic cluster mass function. We have shown how the stellar IMF provides seeds for the growth of massive clumps and have illustrated this with a Salpeter power-law IMF. A more realistic  IMF has a steeper power index at larger stellar masses \citep{kroupa02,chabrier05}. The fragmentation of stellar systems with fewer massive stars would deplete the clump mass function at larger masses more in line with observed statistics for clusters. 
This variability in the clump mass function highlights the major influence of the stellar IMF on the fragmentation process. A full exploration of fragmentation requires hydrodynamical simulations, which we have not performed here. These simulations remain limited to much smaller systems \citep{bate14,lomax14}.

Another physical ingredient that may influence the clump mass function and was not included in the present work is the evolution of massive stars. Significant mass loss by massive stars on a short timescale may induce the dissolution of the larger clumps. It remains to be seen if the inclusion of this effect would transform the bell-shaped function to look more like a power law. Another aspect of dynamical evolution that was not treated in this paper was the inclusion of multiple stars. We have undertaken a project to address these issues and hope to report these findings shortly. In addition, systemic angular momentum is a key difference between the Hubble method we have developed and the fractal approach of \cite{goodwin04}. Angular momentum may be significant in young clusters such as R136 \citep{henault12}. In a fractal model, the way the velocity field is built leaves a residual, global angular momentum whereas the Hubble approach starts off with strictly zero angular momentum.
 A net angular momentum could be introduced in a Hubble model, for instance by setting $\bold{v} = \Hub_o \bold{r} + \bold{\Omega} \times \bold{r}$ with $\bold{\Omega}$ a chosen angular velocity. One can actually go further and write in matrix form~: $\bold{v} = \bold{H}\bold{r}$, with $\bold{H}$ now a 3$\times$3 matrix, where the off-diagonal elements account for rotation and the elements on the diagonal $\Hub_{ii}$ control the three dimensional expansion. In this study, we have set $\Hub_{ij,i\ne j}=0$ and $\Hub_{ii}=\Hub_o$ otherwise. It is then a simple matter to study the fragmentation along a filament by setting, for example, $\Hub_{xx}=\Hub_{yy}=0$ and $\Hub_{zz}>0$. The Hubble fragmentation process for young stellar cluster is a new method and many of its aspects remain to be explored.

\section{Conclusions}

We list here the main findings of this work~:
\begin{itemize}

\item
we introduced a new method to produce sub-virial, substructured initial conditions to explore the dynamical evolution of  young star clusters. Without hydrodynamical calculations and through an Hubble-Lema\^itre like expansion, we induce gravitational fragmentation modes by  adiabatic cooling. Though unrealistic for a full description of cluster formation since it omits, \textit{e.g.}, magnetic field, gas fragmentation and feedback, such a procedure allows the Poissonian fluctuations in the initial density profile to develop over time and yield a self-consistent velocity field and mass distribution~; \\

\item
we tuned  the Minimum Spanning Tree (MST) cut-off method of \cite{maschberger11} to identify the maximum number of clumps in a fragmented configuration (see \S~\ref{Sec:MST}). By doing so,  we eliminate the last free parameter of the method, which allows a more complete comparison of systems with varying degrees of fragmentation~;\\

\item
the clump mass function is sensitive to the stellar IMF. The clump mass function converges more and more towards what is found in hydrodynamical simulations when the upper cut-off mass of a Salpeter IMF is increased (see Fig~\ref{Fig:clumpMF}); \\

\item 
clumps are found to be mass segregated \textit{before} the global collapse and virialisation. This segregation is driven by the formation process of clumps. The mass-segregated clumps bequeath their profiling to the relaxed system (see e.g. \citealt{mcmillan07}). Once virial equilibrium is reached, mass segregation is enhanced  over time on the classic two-body relaxation time-scale~;\\ 

\item 
star clusters that virialised by merging several clumps undergo a softer global infall. Thus their core-halo structure is less concentrated. Their virial  two-body relaxation time is longer than what is obtained from a non-fragmented monolithic collapse of similar initial radius (see \S\S\ref{Sec:collapse} and \ref{Sec:Segregation}). As a result, the merger process leads to virial equilibrium more rapidly. Overall, both monolithic- and fragmented initial conditions lead to clusters with similar mass profiles after some time (about $\sim 1 Myr$ for the case displayed on Fig.~\ref{Fig:Lagrange}), however the fragmented initial conditions leads to a more segregated stellar population;  \\

\item
looking at an out-of-equilibrium system, just after the collapse, the path to equilibrium imprints the spatial distribution of stars differently according to their mass. A by-product of the fragmented Hubble-Lema\^itre formation scenario is that a  broader swat of massive stars segregate to the centre, so enhancing colour gradients in comparison with a formation history proceeding from more uniform, homogeneous mass distributions.


\end{itemize}



\section*{Acknowledgments} 
The work reported here was funded in part by the French national research agency through ANR 2010 JCJC 0501-1 "DESC" (PI E. Moraux). We acknowledge insightful discussions with A. Just, P. Kroupa, E. Vesperini and L. Cambresy. We wish to thank S. Goodwin and the ISSI in Bern (CH) for hosting a workshop   in 2012 when this project was hatched. CMB also thanks M. Gieles for a clever remark concerning collapsing cuspy clusters. Julien Dorval is funded through a PhD grant awarded by the Ecole Doctoral 182 in the University of Strasbourg.


\bibliographystyle{mn2e}
\bibliography{biblio}

\begin{thebibliography}{81}
\expandafter\ifx\csname natexlab\endcsname\relax\def\natexlab#1{#1}\fi

\bibitem[{{Aarseth}(2003)}]{aarseth03}
{Aarseth} S.~J., 2003, {Gravitational N-Body Simulations}. Cambridge University
  Press, Cambridge, UK

\bibitem[{{Aarseth}, {Lin} \& {Papaloizou}(1988){Aarseth}, {Lin}, \&
  {Papaloizou}}]{aarseth88}
{Aarseth} S.~J., {Lin} D.~N.~C., {Papaloizou} J.~C.~B., 1988, \apj, 324, 288

\bibitem[{{Allison} {et~al}\mbox{.}(2009{\natexlab{a}}){Allison}, {Goodwin},
  {Parker}, {de Grijs}, {Portegies Zwart}, \& {Kouwenhoven}}]{allison09}
{Allison} R.~J., {Goodwin} S.~P., {Parker} R.~J., {de Grijs} R., {Portegies
  Zwart} S.~F., {Kouwenhoven} M.~B.~N., 2009{\natexlab{a}}, \apjl, 700, L99

\bibitem[{{Allison} {et~al}\mbox{.}(2010){Allison}, {Goodwin}, {Parker},
  {Portegies Zwart}, \& {de Grijs}}]{allison10}
{Allison} R.~J., {Goodwin} S.~P., {Parker} R.~J., {Portegies Zwart} S.~F., {de
  Grijs} R., 2010, \mnras, 407, 1098

\bibitem[{{Allison} {et~al}\mbox{.}(2009{\natexlab{b}}){Allison}, {Goodwin},
  {Parker}, {Portegies Zwart}, {de Grijs}, \& {Kouwenhoven}}]{allison09b}
{Allison} R.~J., {Goodwin} S.~P., {Parker} R.~J., {Portegies Zwart} S.~F., {de
  Grijs} R., {Kouwenhoven} M.~B.~N., 2009{\natexlab{b}}, \mnras, 395, 1449

\bibitem[{{Andr{\'e}} {et~al}\mbox{.}(2007){Andr{\'e}}, {Belloche}, {Motte}, \&
  {Peretto}}]{andre07}
{Andr{\'e}} P., {Belloche} A., {Motte} F., {Peretto} N., 2007, \aap, 472, 519

\bibitem[{{Andr{\'e}} {et~al}\mbox{.}(2014){Andr{\'e}}, {Di Francesco},
  {Ward-Thompson}, {Inutsuka}, {Pudritz}, \& {Pineda}}]{andre14}
{Andr{\'e}} P., {Di Francesco} J., {Ward-Thompson} D., {Inutsuka} S.-I.,
  {Pudritz} R.~E., {Pineda} J.~E., 2014, in Protostars and Planets VI,
  {Beuther} H., {Klessen} R.~S., {Dullemond} C.~P., {Henning} T.~K., eds., The
  University of Arizona Press, Tucson, AZ, p.~27

\bibitem[{{Andr{\'e}} {et~al}\mbox{.}(2013){Andr{\'e}}, {K{\"o}nyves},
  {Arzoumanian}, {Palmeirim}, \& {Peretto}}]{andre13}
{Andr{\'e}} P., {K{\"o}nyves} V., {Arzoumanian} D., {Palmeirim} P., {Peretto}
  N., 2013, in ASP Conf. Series, Vol. 476, New Trends in Radio Astronomy in the
  ALMA Era, {Kawabe} R., {Kuno} N., {Yamamoto} S., eds., PASP, San Francisco,
  CA, p.~95

\bibitem[{{Bagla} \& {Prasad}(2009)}]{bagla09}
{Bagla} J.~S., {Prasad} J., 2009, \mnras, 393, 607

\bibitem[{{Barnes}, {Lanzel} \& {Williams}(2009){Barnes}, {Lanzel}, \&
  {Williams}}]{barnes09}
{Barnes} E.~I., {Lanzel} P.~A., {Williams} L.~L.~R., 2009, \apj, 704, 372

\bibitem[{{Bastian}, {Covey} \& {Meyer}(2010){Bastian}, {Covey}, \&
  {Meyer}}]{bastian10}
{Bastian} N., {Covey} K.~R., {Meyer} M.~R., 2010, \araa, 48, 339

\bibitem[{{Bate}, {Tricco} \& {Price}(2014){Bate}, {Tricco}, \&
  {Price}}]{bate14}
{Bate} M.~R., {Tricco} T.~S., {Price} D.~J., 2014, \mnras, 437, 77

\bibitem[{{Becker} {et~al}\mbox{.}(2013){Becker}, {Moraux}, {Duch{\^e}ne},
  {Maschberger}, \& {Lawson}}]{becker13}
{Becker} C., {Moraux} E., {Duch{\^e}ne} G., {Maschberger} T., {Lawson} W.,
  2013, \aap, 552, A46

\bibitem[{{Benhaiem} \& {Sylos Labini}(2015)}]{benhaiem15}
{Benhaiem} D., {Sylos Labini} F., 2015, \mnras, 448, 2634

\bibitem[{{Binney} \& {Tremaine}(2008)}]{BT08}
{Binney} J., {Tremaine} S., 2008, {Galactic Dynamics (2nd ed.)}. Princeton
  Univ. Press, Princeton, NJ

\bibitem[{{Boily}, {Athanassoula} \& {Kroupa}(2002){Boily}, {Athanassoula}, \&
  {Kroupa}}]{boily02}
{Boily} C.~M., {Athanassoula} E., {Kroupa} P., 2002, \mnras, 332, 971

\bibitem[{{Boily}, {Clarke} \& {Murray}(1999){Boily}, {Clarke}, \&
  {Murray}}]{boily99}
{Boily} C.~M., {Clarke} C.~J., {Murray} S.~D., 1999, \mnras, 302, 399

\bibitem[{{Bonnell}, {Vine} \& {Bate}(2004){Bonnell}, {Vine}, \&
  {Bate}}]{bonnell04}
{Bonnell} I.~A., {Vine} S.~G., {Bate} M.~R., 2004, \mnras, 349, 735

\bibitem[{{Cambr{\'e}sy} {et~al}\mbox{.}(2006){Cambr{\'e}sy}, {Petropoulou},
  {Kontizas}, \& {Kontizas}}]{cambresy06}
{Cambr{\'e}sy} L., {Petropoulou} V., {Kontizas} M., {Kontizas} E., 2006, \aap,
  445, 999

\bibitem[{{Caputo}, {de Vries} \& {Portegies Zwart}(2014){Caputo}, {de Vries},
  \& {Portegies Zwart}}]{caputo14}
{Caputo} D.~P., {de Vries} N., {Portegies Zwart} S., 2014, \mnras, 445, 674

\bibitem[{{Chabrier}(2005)}]{chabrier05}
{Chabrier} G., 2005, in Astr. \& Space Sci., Vol. 327, The Initial Mass
  Function 50 Years Later, {Corbelli} E., {Palla} F., {Zinnecker} H., eds.,
  p.~41

\bibitem[{{Cortes} {et~al}\mbox{.}(2010){Cortes}, {Parra}, {Cortes}, \&
  {Hardy}}]{cortes10}
{Cortes} P.~C., {Parra} R., {Cortes} J.~R., {Hardy} E., 2010, \aap, 519, A35

\bibitem[{{Cottaar} {et~al}\mbox{.}(2015){Cottaar}, {Covey}, {Foster}, {Meyer},
  {Tan}, {Nidever}, {Chojnowski}, {da Rio}, {Flaherty}, {Frinchaboy},
  {Majewski}, {Skrutskie}, {Wilson}, \& {Zasowski}}]{cottaar15}
{Cottaar} M. {et~al.}, 2015, \apj, 807, 27

\bibitem[{{Czekaj} {et~al}\mbox{.}(2014){Czekaj}, {Robin}, {Figueras}, {Luri},
  \& {Haywood}}]{czekaj14}
{Czekaj} M.~A., {Robin} A.~C., {Figueras} F., {Luri} X., {Haywood} M., 2014,
  \aap, 564, A102

\bibitem[{{Ehlerova} {et~al}\mbox{.}(1997){Ehlerova}, {Palous}, {Theis}, \&
  {Hensler}}]{ehlerova97}
{Ehlerova} S., {Palous} J., {Theis} C., {Hensler} G., 1997, \aap, 328, 121

\bibitem[{{Eisenstein} \& {Hut}(1998)}]{eisenstein98}
{Eisenstein} D.~J., {Hut} P., 1998, \apj, 498, 137

\bibitem[{{Fleck} {et~al}\mbox{.}(2006){Fleck}, {Boily}, {Lan{\c c}on}, \&
  {Deiters}}]{fleck06}
{Fleck} J.-J., {Boily} C.~M., {Lan{\c c}on} A., {Deiters} S., 2006, \mnras,
  369, 1392

\bibitem[{{Foster} {et~al}\mbox{.}(2015){Foster}, {Cottaar}, {Covey}, {Arce},
  {Meyer}, {Nidever}, {Stassun}, {Tan}, {Chojnowski}, {da Rio}, {Flaherty},
  {Rebull}, {Frinchaboy}, {Majewski}, {Skrutskie}, {Wilson}, \&
  {Zasowski}}]{foster15}
{Foster} J.~B. {et~al.}, 2015, \apj, 799, 136

\bibitem[{Freeman {et~al}\mbox{.}(2015)Freeman, Block, Elmegreen, \&
  Woolway}]{freeman15}
Freeman K.~C., Block D., Elmegreen B.~G., Woolway M., 2015, Lessons from the
  local group : a conference in honour of David Block and Bruce Elmegreen.
  Springer

\bibitem[{{Friedman} \& {Schutz}(1978)}]{friedman78}
{Friedman} J.~L., {Schutz} B.~F., 1978, \apj, 221, 937

\bibitem[{{Fujii} \& {Portegies Zwart}(2016)}]{fujii16}
{Fujii} M.~S., {Portegies Zwart} S., 2016, \apj, 817, 4

\bibitem[{{Fujii}, {Saitoh} \& {Portegies Zwart}(2012){Fujii}, {Saitoh}, \&
  {Portegies Zwart}}]{fujii12}
{Fujii} M.~S., {Saitoh} T.~R., {Portegies Zwart} S.~F., 2012, \apj, 753, 85

\bibitem[{{Goodwin} \& {Whitworth}(2004)}]{goodwin04}
{Goodwin} S.~P., {Whitworth} A.~P., 2004, \aap, 413, 929

\bibitem[{{Gunawardhana} {et~al}\mbox{.}(2011){Gunawardhana}, {Hopkins},
  {Sharp}, {Brough}, {Taylor}, {Bland-Hawthorn}, {Maraston}, {Tuffs},
  {Popescu}, {Wijesinghe}, {Jones}, {Croom}, {Sadler}, {Wilkins}, {Driver},
  {Liske}, {Norberg}, {Baldry}, {Bamford}, {Loveday}, {Peacock}, {Robotham},
  {Zucker}, {Parker}, {Conselice}, {Cameron}, {Frenk}, {Hill}, {Kelvin},
  {Kuijken}, {Madore}, {Nichol}, {Parkinson}, {Pimbblet}, {Prescott},
  {Sutherland}, {Thomas}, \& {van Kampen}}]{gunawardhana11}
{Gunawardhana} M.~L.~P. {et~al.}, 2011, \mnras, 415, 1647

\bibitem[{{Gutermuth} {et~al}\mbox{.}(2009){Gutermuth}, {Megeath}, {Myers},
  {Allen}, {Pipher}, \& {Fazio}}]{gutermuth09}
{Gutermuth} R.~A., {Megeath} S.~T., {Myers} P.~C., {Allen} L.~E., {Pipher}
  J.~L., {Fazio} G.~G., 2009, \apjs, 184, 18

\bibitem[{{Haas} \& {Anders}(2010)}]{haas10}
{Haas} M.~R., {Anders} P., 2010, \aap, 512, A79

\bibitem[{{Haghi} {et~al}\mbox{.}(2014){Haghi}, {Hoseini-Rad}, {Zonoozi}, \&
  {K{\"u}pper}}]{haghi14}
{Haghi} H., {Hoseini-Rad} S.~M., {Zonoozi} A.~H., {K{\"u}pper} A.~H.~W., 2014,
  \mnras, 444, 3699

\bibitem[{{Heggie} \& {Mathieu}(1986)}]{heggie86}
{Heggie} D.~C., {Mathieu} R.~D., 1986, in Lecture Notes in Physics, Vol. 267,
  The Use of Supercomputers in Stellar Dynamics, {Hut} P., {McMillan} S.~L.~W.,
  eds., Springer Verlag, Berlin, p. 233

\bibitem[{{H{\'e}nault-Brunet} {et~al}\mbox{.}(2012){H{\'e}nault-Brunet},
  {Gieles}, {Evans}, {Sana}, {Bastian}, {Ma{\'{\i}}z Apell{\'a}niz}, {Taylor},
  {Markova}, {Bressert}, {de Koter}, \& {van Loon}}]{henault12}
{H{\'e}nault-Brunet} V. {et~al.}, 2012, \aap, 545, L1

\bibitem[{{H\'enon}(1973)}]{henon73}
{H\'enon} M., 1973, \aap, 24, 229

\bibitem[{{Hillenbrand} \& {Hartmann}(1998)}]{hillenbrand98}
{Hillenbrand} L.~A., {Hartmann} L.~W., 1998, \apj, 492, 540

\bibitem[{{Hoversten} \& {Glazebrook}(2008)}]{hoversten08}
{Hoversten} E.~A., {Glazebrook} K., 2008, \apj, 675, 163

\bibitem[{{Kirk} \& {Myers}(2011)}]{kirk11}
{Kirk} H., {Myers} P.~C., 2011, \apj, 727, 64

\bibitem[{{Klessen} \& {Burkert}(2000)}]{klessen00}
{Klessen} R.~S., {Burkert} A., 2000, \apjs, 128, 287

\bibitem[{{Klessen} \& {Burkert}(2001)}]{klessen01}
{Klessen} R.~S., {Burkert} A., 2001, \apj, 549, 386

\bibitem[{{Kroupa}(2002)}]{kroupa02}
{Kroupa} P., 2002, Science, 295, 82

\bibitem[{{K{\"u}pper} {et~al}\mbox{.}(2011){K{\"u}pper}, {Maschberger},
  {Kroupa}, \& {Baumgardt}}]{kupper11}
{K{\"u}pper} A.~H.~W., {Maschberger} T., {Kroupa} P., {Baumgardt} H., 2011,
  \mnras, 417, 2300

\bibitem[{{Lada} \& {Lada}(2003)}]{ladalada03}
{Lada} C.~J., {Lada} E.~A., 2003, \araa, 41, 57

\bibitem[{{Lomax} {et~al}\mbox{.}(2014){Lomax}, {Whitworth}, {Hubber},
  {Stamatellos}, \& {Walch}}]{lomax14}
{Lomax} O., {Whitworth} A.~P., {Hubber} D.~A., {Stamatellos} D., {Walch} S.,
  2014, \mnras, 439, 3039

\bibitem[{{Mac Low} \& {Klessen}(2004)}]{maclow04}
{Mac Low} M.-M., {Klessen} R.~S., 2004, Rev. Mod. Phys., 76, 125

\bibitem[{{Marks} \& {Kroupa}(2012)}]{marks12}
{Marks} M., {Kroupa} P., 2012, \aap, 543, A8

\bibitem[{{Maschberger} \& {Clarke}(2011)}]{maschberger11}
{Maschberger} T., {Clarke} C.~J., 2011, \mnras, 416, 541

\bibitem[{{Maschberger} {et~al}\mbox{.}(2010){Maschberger}, {Clarke},
  {Bonnell}, \& {Kroupa}}]{maschberger10}
{Maschberger} T., {Clarke} C.~J., {Bonnell} I.~A., {Kroupa} P., 2010, \mnras,
  404, 1061

\bibitem[{{McGlynn}(1984)}]{mcglynn84}
{McGlynn} T.~A., 1984, \apj, 281, 13

\bibitem[{{McMillan}, {Vesperini} \& {Portegies Zwart}(2007){McMillan},
  {Vesperini}, \& {Portegies Zwart}}]{mcmillan07}
{McMillan} S.~L.~W., {Vesperini} E., {Portegies Zwart} S.~F., 2007, \apjl, 655,
  L45

\bibitem[{{Merritt}(2013)}]{merritt13}
{Merritt} D., 2013, {Dynamics and Evolution of Galactic Nuclei}. Princeton
  Univ. Press, Princeton, NJ

\bibitem[{{Meylan} \& {Heggie}(1997)}]{meylan97}
{Meylan} G., {Heggie} D.~C., 1997, \aapr, 8, 1

\bibitem[{{Moeckel} \& {Clarke}(2011)}]{moeckel11}
{Moeckel} N., {Clarke} C.~J., 2011, \mnras, 410, 2799

\bibitem[{{Moeckel} {et~al}\mbox{.}(2012){Moeckel}, {Holland}, {Clarke}, \&
  {Bonnell}}]{moeckel12}
{Moeckel} N., {Holland} C., {Clarke} C.~J., {Bonnell} I.~A., 2012, \mnras, 425,
  450

\bibitem[{{Moraux} {et~al}\mbox{.}(2007){Moraux}, {Bouvier}, {Stauffer},
  {Barrado y Navascu{\'e}s}, \& {Cuillandre}}]{moraux07}
{Moraux} E., {Bouvier} J., {Stauffer} J.~R., {Barrado y Navascu{\'e}s} D.,
  {Cuillandre} J.-C., 2007, \aap, 471, 499

\bibitem[{{Moraux} {et~al}\mbox{.}(2003){Moraux}, {Bouvier}, {Stauffer}, \&
  {Cuillandre}}]{moraux03}
{Moraux} E., {Bouvier} J., {Stauffer} J.~R., {Cuillandre} J.-C., 2003, \aap,
  400, 891

\bibitem[{{Motte}, {Schilke} \& {Lis}(2003){Motte}, {Schilke}, \&
  {Lis}}]{motte03}
{Motte} F., {Schilke} P., {Lis} D.~C., 2003, \apj, 582, 277

\bibitem[{{Nitadori} \& {Aarseth}(2012)}]{nitadori12}
{Nitadori} K., {Aarseth} S.~J., 2012, \mnras, 424, 545

\bibitem[{{Offner} {et~al}\mbox{.}(2014){Offner}, {Clark}, {Hennebelle},
  {Bastian}, {Bate}, {Hopkins}, {Moraux}, \& {Whitworth}}]{offner14}
{Offner} S.~S.~R., {Clark} P.~C., {Hennebelle} P., {Bastian} N., {Bate} M.~R.,
  {Hopkins} P.~F., {Moraux} E., {Whitworth} A.~P., 2014, in Protostars and
  Planets VI, {Beuther} H., {Klessen} R.~S., {Dullemond} C.~P., {Henning}
  T.~K., eds., The University of Arizona Press, Tucson, AZ, p.~53

\bibitem[{{Olczak}, {Spurzem} \& {Henning}(2011){Olczak}, {Spurzem}, \&
  {Henning}}]{olczak11}
{Olczak} C., {Spurzem} R., {Henning} T., 2011, \aap, 532, A119

\bibitem[{{Parker}, {Goodwin} \& {Allison}(2011){Parker}, {Goodwin}, \&
  {Allison}}]{parker11}
{Parker} R.~J., {Goodwin} S.~P., {Allison} R.~J., 2011, \mnras, 418, 2565

\bibitem[{{Parker} {et~al}\mbox{.}(2014){Parker}, {Wright}, {Goodwin}, \&
  {Meyer}}]{parker14}
{Parker} R.~J., {Wright} N.~J., {Goodwin} S.~P., {Meyer} M.~R., 2014, \mnras,
  438, 620

\bibitem[{{Peebles}(1980)}]{peebles80}
{Peebles} P.~J.~E., 1980, {The large-scale structure of the universe}.
  Princeton Univ. Press, Princeton, NJ

\bibitem[{{Portegies Zwart}, {McMillan} \& {Gieles}(2010){Portegies Zwart},
  {McMillan}, \& {Gieles}}]{SPZ10}
{Portegies Zwart} S.~F., {McMillan} S.~L.~W., {Gieles} M., 2010, \araa, 48, 431

\bibitem[{{Rathborne} {et~al}\mbox{.}(2015){Rathborne}, {Longmore}, {Jackson},
  {Alves}, {Bally}, {Bastian}, {Contreras}, {Foster}, {Garay}, {Kruijssen},
  {Testi}, \& {Walsh}}]{rathborne15}
{Rathborne} J.~M. {et~al.}, 2015, \apj, 802, 125

\bibitem[{{Reipurth} \& {Clarke}(2001)}]{reipurth01}
{Reipurth} B., {Clarke} C., 2001, \aj, 122, 432

\bibitem[{{Reipurth} {et~al}\mbox{.}(2014){Reipurth}, {Clarke}, {Boss},
  {Goodwin}, {Rodr{\'{\i}}guez}, {Stassun}, {Tokovinin}, \&
  {Zinnecker}}]{reipurth14}
{Reipurth} B., {Clarke} C.~J., {Boss} A.~P., {Goodwin} S.~P.,
  {Rodr{\'{\i}}guez} L.~F., {Stassun} K.~G., {Tokovinin} A., {Zinnecker} H.,
  2014, in Protostars and Planets VI, {Beuther} H., {Klessen} R.~S.,
  {Dullemond} C.~P., {Henning} T.~K., eds., The University of Arizona Press,
  Tucson, AZ, p. 267

\bibitem[{{Rybizki} \& {Just}(2015)}]{rybizki15}
{Rybizki} J., {Just} A., 2015, \mnras, 447, 3880

\bibitem[{{Salpeter}(1955)}]{salpeter55}
{Salpeter} E.~E., 1955, \apj, 121, 161

\bibitem[{{Skory} {et~al}\mbox{.}(2010){Skory}, {Turk}, {Norman}, \&
  {Coil}}]{skory10}
{Skory} S., {Turk} M.~J., {Norman} M.~L., {Coil} A.~L., 2010, \apjs, 191, 43

\bibitem[{{Theis} \& {Spurzem}(1999)}]{theis99}
{Theis} C., {Spurzem} R., 1999, \aap, 341, 361

\bibitem[{{van Albada}(1982)}]{vanalbada82}
{van Albada} T.~S., 1982, \mnras, 201, 939

\bibitem[{{Vesperini} {et~al}\mbox{.}(2014){Vesperini}, {Varri}, {McMillan}, \&
  {Zepf}}]{vesperini14}
{Vesperini} E., {Varri} A.~L., {McMillan} S.~L.~W., {Zepf} S.~E., 2014, \mnras,
  443, L79

\bibitem[{{Weidner} \& {Kroupa}(2005)}]{weidner05}
{Weidner} C., {Kroupa} P., 2005, \apj, 625, 754

\bibitem[{{Weidner}, {Kroupa} \& {Pflamm-Altenburg}(2013){Weidner}, {Kroupa},
  \& {Pflamm-Altenburg}}]{weidner13}
{Weidner} C., {Kroupa} P., {Pflamm-Altenburg} J., 2013, \mnras, 434, 84

\bibitem[{{W{\"u}nsch} {et~al}\mbox{.}(2010){W{\"u}nsch}, {Dale}, {Palou{\v
  s}}, \& {Whitworth}}]{wunsch10}
{W{\"u}nsch} R., {Dale} J.~E., {Palou{\v s}} J., {Whitworth} A.~P., 2010,
  \mnras, 407, 1963

\end{thebibliography}


\newpage 
\appendix

\section{Identifying bound stars}
\begin{figure}
\begin{center}
\includegraphics[width=\columnwidth]{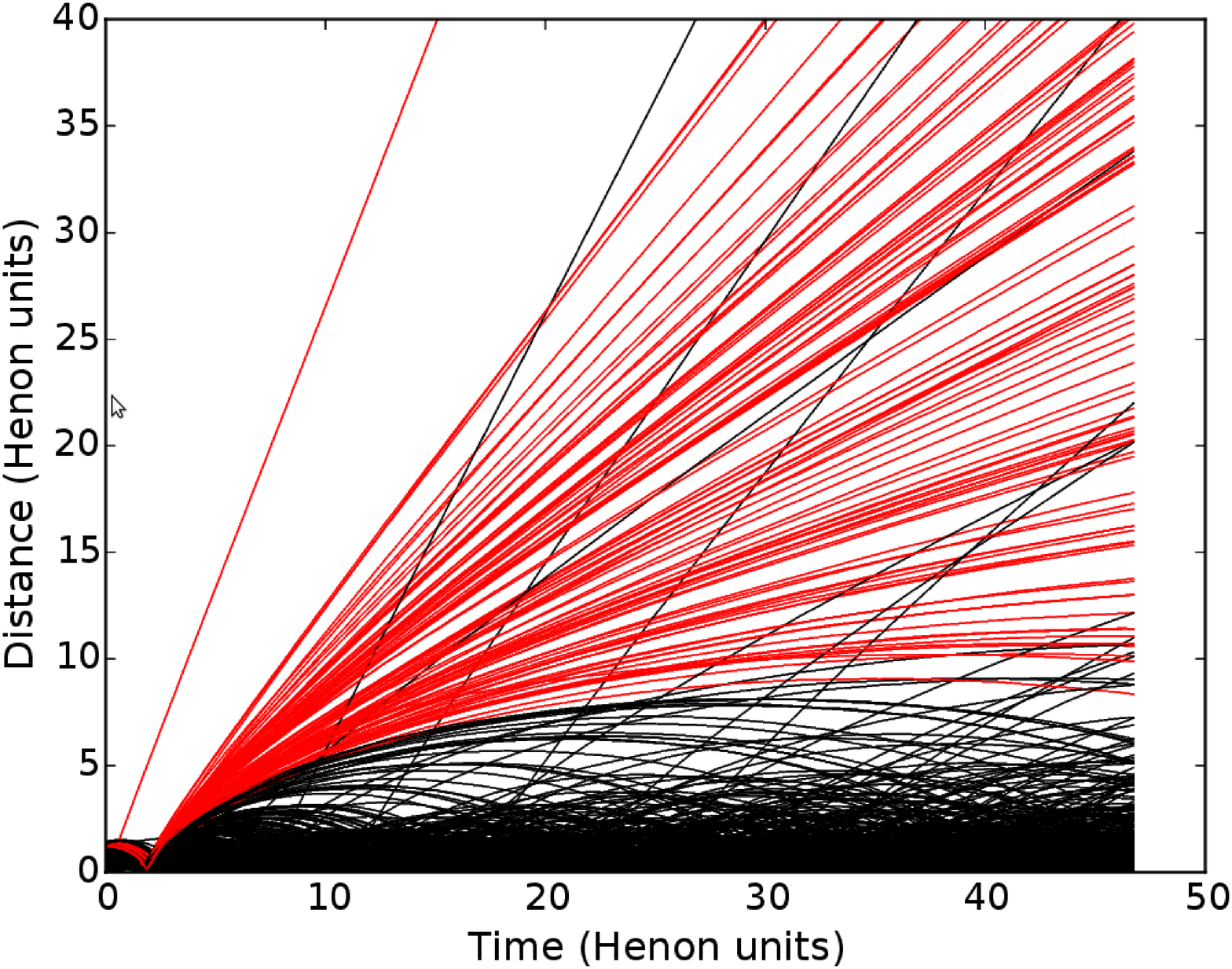}
\caption{Distance to origin for 750 stars from run Ru20 (see Table~\ref{Tab:evolution_models}). Red lines show the trajectory of stars that are considered ejected according to our criterion.}
\label{Fig:DistOrigin}
\end{center}
\end{figure}

To understand better the evolution of the bound stars only, we needed to isolate and substract the ejected stars from the system as a whole. The obvious way to do this would be to compute the stars mechanical energy and to remove all stars with positive energy. Though this works for a majority of the ejected stars, a subset of them has a marginally negative energy. These register as bound when they are essentially out of the system (far beyond the original system radius) and do not contribute to the dynamics. To collect a maximum number of ejected stars efficiently, we spotted the time when the potential energy is maximum, when the collapse occurs. We  then identified all stars whose distance to the center increased monotonically from there onwards. The full selection criteria is therefore~: $v_r(t) > 0, E_\star > 0 \, \forall t > t_{ff}$. 
This allows a more complete selection of the ejecta.  On Fig.~\ref{Fig:DistOrigin} we graph  $|\bold{r}|$  as a function of time for a subset of escapers (shown as red curves) for the uniform collapse model Ru20. The black curves are trajectories for bound stars given for comparison. Some of these bound stars are later ejected from the system due to two-body interactions, as seen on the figure.

\section{Time of the end of expansion}\label{App:Hubble}

In this section we detail the derivation of the analytical expression for the time at which the expansion stops and the system starts to collapse on itself. We start from a uniform sphere of radius $R_0$, total mass $M$. We consider spherical shells as mass elements, situated at distance $r$ from the origin. They are attributed a radial velocity following (for the shell at $r=R_0$) $\vec v_0 = \Hub_0 \vec R_0 = \Hub_0 R_0 \vec u_r$ with $\Hub_0$ being a parameter akin to the Hubble constant in cosmology. We want to follow the radial motion of the last shell of mass m, situated at $R$ from the origin. Newton's second law gives:

\begin{align}\label{eq:newton}
m \frac{dv}{dt} & = - \frac{G M m}{R^2}
\end{align}

By multiplying on both sides by $v$ and integrating between a given time and $t=0$, one finds:

\begin{equation}
v^2(t) - v^2_0 = 2GM \left( \inv{R} - \inv{R_0} \right)
\end{equation}%

Which becomes, by taking $\nu = v/v0$,  $x= R/R0$ and $E_\ast = \frac{2GM}{R_0 v_0^2}$, which is a dimensionless measure of the total energy of the system:

\begin{equation}
\nu^2  = 1 + E_\ast \left( \inv{x} - 1 \right) .
\end{equation}

The evolution of the system has 3 outcomes, depending on the value of $E_\ast$:
\begin{itemize}
\item $E_\ast<1$ The velocity is always strictly positive as the system expands ($x->\infty$). The system is unbound.
\item $E_\ast=1$ The velocity approaches zero as the system expands. The expansion "stops at an infinite radius". The system is marginally bound.
\item $E_\ast>1$ The velocity reaches zero for a finite radius, the system is bound and will collapses back on itself once the expansion stops. 
\end{itemize}

We only consider in the following the case in which $E_\ast<1$. We have the expression

\begin{equation}
\nu = \sqrt{1+E_\ast\left(\inv{x} - 1\right)}
\end{equation}

Taking the time derivative gives:

\begin{equation}
\frac{d \nu}{dt} = - \frac{E_\ast}{2 x^2} \left[ 1 + E_\ast\left(\inv{x} -1\right)\right]^{-\frac{1}{2}} \frac{dx}{dt}
\end{equation}

Combining this with (\ref{eq:newton}), one obtains:

\begin{equation}
\frac{dx}{dt} = \Hub_0 \sqrt{1+ E_\ast\left( \inv{x} -1\right)}
\end{equation}

which can be rewritten, using $\tilde{\Hub_0} = \Hub_0 \sqrt{E_\ast-1}$ and $x_t=\frac{E_\ast}{E_\ast-1}$

\begin{equation}
\frac{dx}{dt} = \tilde{\Hub_0} \sqrt{\frac{x_t}{x}-1}
\end{equation}

$x_t$ being the extent of the maximum expansion as we assumed a bound system. The subscript t is for "turn-around". If we choose the notation $u = \frac{x}{x_t}$:

\begin{equation}
\sqrt{\frac{u}{u-1}} \frac{du}{dt} = \frac{\tilde{\Hub_0}}{x_t}
\end{equation}

We know that $x$ varies from 1 to $x_t$, thus $u$ varies from $1/x_t$ to 1. We can then make the change of variable $u = \sin^2\theta$ and separate the variables:

\begin{equation}
\sqrt{\frac{\sin^2\theta}{1-\sin^2\theta}} 2 \sin\theta \cos\theta d \theta = \frac{\tilde{\Hub_0}}{x_t} dt
\end{equation}

which becomes after simplifications:

\begin{equation}
[ 1 - \cos(2\theta)]d\theta = \frac{\tilde{\Hub_0}}{x_t} dt .
\end{equation}

We now integrate the expression from $t=0$ to $t$, the time at which the expansions stops and $x$ reaches $x_t$ (wich implies $u_t = 1$ and $\theta_t = \pi /2)$:

\begin{align}
\int^{\pi/2}_{\theta_0} [ 1 - \cos(2\theta)]d\theta  & = \int^t_0 \frac{\tilde{\Hub_0}}{x_t} dt\\
\frac{\pi}{2} - \theta_0 + \frac{\sin(2\theta_0)}{2} & =  \frac{\tilde{\Hub_0}}{x_t} t\\
\pi - 2 \theta_0 + \frac{2}{\sqrt{x_t}}\sqrt{1-\inv{x_t}} & = 2 \frac{\tilde{\Hub_0}}{x_t} t 
\end{align}

which boils down to the expression of the time at which the expansion stops:

\begin{equation}
t = \frac{E_\ast \left(\frac{\pi}{2} - \theta_0\right) + \sqrt{E_\ast-1}}{\Hub_0 (E_\ast-1)^{-\frac{3}{2}}}.
\end{equation}

Recalling the quantities:
\begin{align}
E_\ast = \frac{2GM}{R_0 v_0^2}        &;  & x_t=\frac{E_\ast}{E_\ast-1}  &;  &\theta_0 = \sin^{-1}\left(\inv{\sqrt{x_t}}\right)  
\end{align}

\section{Power spectrum}

In section \ref{sec:velocityfield}, we mentionned the hydrodynamical simulations of \cite{klessen00}. In such simulations, the initial power spectrum is of major importance. In this appendix, we take a look at the initial power spectrum both in theirs and our models. \cite{klessen00} had their SPH simulation run for a system free-fall time starting from a smooth (isothermal)  velocity field,  but a perturbed  density profile. The perturbations took the form of a random Gaussian field, with power spectrum $P_k \propto k^{-2}$, i.e.  a power-law of the sampled wavenumber $k$. Similar results are given in a follow-up study by \cite{klessen01}, where the spectral index $k$ was varied between 0 and -3. \citeauthor{klessen01} noted that the characteristics of their star-forming clumps were not much different despite the broad  range of initial density perturbations probed. The weak influence on the shape of the power spectrum can be understood in the light of work by \cite{bagla09}, who showed that the small-scale structuration is not strongly affected by the large scale modes. This prompted us to ask in what respect the result obtained from Poisson fluctuations, which seeded our density profiles, differ from those derived from Gaussian random fields. 

To answer that question, we recast our initial problem of \S \ref{Sec:FourierModes} of sampling a uniform density profile with $N$ mass elements in terms of the  mass spectrum of the stars. We can then write for the mass density 
\[ \rho = N_v \, \langle m_\star\rangle \] 
where $N_v$ denotes the expected number of stars within volume $V$, and $< .. >$ denotes number-averaging. If we sample $V$ and assume some statistical noise, then 

\begin{equation} \label{Eqn:Densityfluctuations} 
\left<\frac{\delta\rho}{\rho}\right> = \left< \frac{\delta N_v}{N_v} \right> + \left< \frac{\delta \langle m_\star\rangle}{\langle m_\star\rangle} \right> \ .  
\end{equation} 

This equation shows that density fluctuations will be minimal when the mass spectrum is narrow, with equal-mass  stars yielding the absolute minimum. With the stars initial positions being uncorrelated with their mass, the statistical noise will be Poissonian for both the expectations for $N_v$ {\it  and} mean stellar mass. We can relate the volume being sampled by a Fourier mode of wavenumber $k = 2\pi / \lambda$ by  $V = 4\pi ( \lambda/2)^3/3 \simeq 4 \pi^3 k^{-3}$. Equation (\ref{Eqn:Densityfluctuations}) is rewritten in the general form as

\begin{equation} \label{Eqn:Densityfluctuations2} 
\left<\frac{\delta\rho}{\rho}\right> = ( 1 + C_\alpha) \left< \frac{\delta N_v}{N_v} \right> \simeq 
( 1 + C_\alpha ) \left(\frac{V}{4\pi^3N}\right)^{1/2} k^{3/2} \, . 
\end{equation} 

In the last equation, we have introduced a proportionality constant $C_\alpha > 0 $ that will change with the chosen stellar mass function ($\alpha$ being the Salpeter index).  The power spectrum defined as the averaged squared amplitudes $\langle (\delta\rho/\rho)^2\rangle $ leads to $P(k) \propto k^{3}$, so that the power spectrum peaks at short wavelengths $\lambda$. Therefore in N-body calculations of fragmentation modes, small clumps form first and are constantly merging with one another, forming larger associations  hierarchically (the `bottom-up' picture of cosmology). The setup of \cite{klessen00,klessen01} takes  the opposite 'top-down' route. It also interesting to note that the bounds given to the mass function only affect the amplitude of the power spectrum, not its shape.

\end{document}